\documentclass[]{aastex}
\usepackage{emulateapj5}
\usepackage{onecolfloat}
\usepackage{graphicx} 
\usepackage{fancyheadings} 
\usepackage{ulem}
\usepackage{rotating}
\usepackage{lscape}

\newcommand{\nfin}{19}
\newcommand{\nnew}{14}
\newcommand{\nstud}{six}
\newcommand{\nlit}{five}
\newcommand{\suii}{S$^+$}
\newcommand{\suiii}{S$^{++}$}
\newcommand{\hi}{H$^0$}

\newcommand{\oi}{O$^0$}
\newcommand{\nti}{N$^0$}
\newcommand{\ntii}{N$^+$}
\newcommand{\ari}{Ar$^0$}

\newcommand{\alii}{Al$^+$}
\newcommand{\aliii}{Al$^{++}$}
\newcommand{\siii}{Si$^+$}
\newcommand{\siiii}{Si$^{++}$}

\newcommand{\feii}{Fe$^+$}
\newcommand{\feiii}{Fe$^{++}$}

\newcommand{\nalph}{N/$\alpha$}
\newcommand{\alphH}{$\alpha$/H}

\newcommand{\msol}{M_\odot}

\newcommand{\lya}{Ly$\alpha$ }

\newcommand{\cm}[1]{\, {\rm cm^{#1}}}
\newcommand{\N}[1]{{N({\rm #1})}}

\newcommand{\sci}[1]{{\rm \; \times \; 10^{#1}}}

\newcommand{\mkms}{{\rm \; km\;s^{-1}}}

\begin{document}

\twocolumn[%
\submitted{Accepted to the Astrophysical Journal: June 14, 2002}

\title{THE UCSD HIRES/KECK I DAMPED \lya ABUNDANCE 
DATABASE:\altaffilmark{1}
IV. Probing Galactic Enrichment Histories with Nitrogen}

\author{ JASON X. PROCHASKA\altaffilmark{2,3}}
\affil{The Observatories of the Carnegie Institute of Washington}
\affil{813 Santa Barbara St. \\
Pasadena, CA 91101}
\email{xavier@ociw.edu}
\and
\author{ RICHARD B.C. HENRY}
\affil{Department of Physics and Astronomy;
University of Oklahoma; 
Norman, OK 73019}
\email{henry@mail.nhn.ou.edu}
\and
\author{JOHN M. O'MEARA\altaffilmark{2}, DAVID TYTLER\altaffilmark{2},
 ARTHUR M. WOLFE\altaffilmark{2}, DAVID KIRKMAN\altaffilmark{2},
DAN LUBIN, \& NAO SUZUKI}
\affil{Department of Physics, and Center for Astrophysics and Space Sciences}
\affil{University of California, San Diego; 
C--0424; La Jolla, CA 92093}
\email{jomeara@ucsd.edu, dtytler@ucsd.edu, awolfe@ucsd.edu, dkirkman@ucsd.edu
dlubin@ucsd.edu, n1suzuki@ucsd.edu}

\begin{abstract} 

We present \nnew\ N$^0$ measurements from our HIRES/Keck database of
damped \lya abundances.  These data are combined with measurements from the 
recent and past literature to build an homogeneous, uniform set of observations.
We examine photoionization diagnostics like Fe$^{++}$ and Ar$^0$ 
in the majority of the complete sample and assess the impact 
of ionization corrections on \nalph\ and \alphH\ values derived 
from observed ionic column densities of N$^0$, 
Si$^+$, H$^0$, and S$^+$.  Our final sample of \nfin\ \nalph, \alphH\ pairs
appears bimodal; the majority of systems show \nalph\ values consistent with
metal-poor emission regions in the local universe but a small sub-sample
exhibit significantly lower \nalph\ ratios.
Contrary to previous studies of \nalph\ in the damped systems, our sample
shows little scatter within each sub-sample.
We consider various scenarios to explain the presence 
of the low \nalph\ sightlines
and account for the apparent bimodality.  We favor a model where at least
some galaxies undergo an initial burst of star formation with 
suppressed formation of intermediate-mass stars.  
We found a power-law IMF with slope 0.10 or a 
mass cut of $\approx 5-8 \msol$ would successfully reproduce 
the observed LN-DLA values.
If the bimodal distribution is confirmed by a larger sample of
measurements,
this may present the first observational evidence for a top 
heavy initial mass function in some early stellar populations. 

\end{abstract}

\keywords{galaxies: abundances --- 
galaxies: chemical evolution --- quasars : absorption lines ---
nucleosynthesis}
]

\pagestyle{fancyplain}
\lhead[\fancyplain{}{\thepage}]{\fancyplain{}{PROCHASKA ET AL.}}
\rhead[\fancyplain{}{THE UCSD HIRES/KECK\,I DAMPED \lya ABUNDANCE 
DATABASE IV.}]{\fancyplain{}{\thepage}}
\setlength{\headrulewidth=0pt}
\cfoot{}

\altaffiltext{1}{http://kingpin.ucsd.edu/$\sim$hiresdla} 
\altaffiltext{2}{Visiting Astronomer, W.M. Keck Telescope.
The Keck Observatory is a joint facility of the University
of California and the California Institute of Technology.}
\altaffiltext{3}{Hubble fellow}

\section{INTRODUCTION}
\label{sec:intro}

Abundance studies of the damped \lya systems -- protogalaxies with
large H\,I surface densities $\N{HI} > 2 \sci{20} \cm{-2}$ --
reveal the chemical enrichment history of the universe.
These quasar absorption line systems dominate the neutral hydrogen
reservoir to $z \sim 4$ \citep{wol95,storr00,peroux01}
and, therefore, their metal abundances presumably reflect the 
past and current processes of protogalactic
star formation \citep[e.g.][]{pei99}.  High resolution
surveys of the damped systems measure the metal content of these
galaxies and trace the evolution in Zn and
Fe metallicity with redshift \citep{ptt94,pw00}.  Although these 
observations track the gross census of metals with time, they 
only crudely describe the physical processes 
of metal enrichment.  To examine issues related to the initial mass
function (IMF), the formation epoch, and the star formation rate of
individual systems, one must pursue other diagnostics.

One important avenue for addressing the details of chemical enrichment
is through the investigation of relative metal abundances like O/Fe
\citep[e.g.][]{tinsley79}.  By comparing the so-called 
$\alpha$-elements, elements presumed to
form primarily in massive star supernovae, 
against Fe one roughly assesses the relative contribution of Type~Ia
and Type~II SN which relates to the IMF and star formation history.
Although current observations of the damped systems \citep{pw02}
suggest an $\alpha$-enhancement at low metallicity
similar to Galactic metal-poor stars \citep{wheeler89,mcw97}, these
measurements are subject to the effects of differential depletion
\citep{vladilo98,ledoux02}.  
To date, it has proved a great challenge to isolate the competing
effects of nucleosynthesis and differential depletion for ratios like Si/Fe
and O/Fe.
One can minimize these uncertainties 
by examining a larger set of $\alpha$ and Fe-peak
elements \citep[e.g.][]{dessauges02a},
but this requires extensive observations and is still subject to some
uncertainty. 

A complementary approach toward tracing the detailed enrichment history
of protogalaxies is to consider
the relative abundance of nitrogen. Nitrogen is mainly produced in 
the six steps of the CN branch of the
CNO cycles within H burning stellar zones, where $^{12}$C serves as
the reaction catalyst (see a textbook like Clayton 1983 or Cowley 1995
for a nucleosynthesis review).  Three reactions occur to transform
$^{12}$C to $^{14}$N: $^{12}$C(p,$\gamma$)$^{13}$N($\beta$$^{+}$,$
\nu$)$^{13}$C(p,$\gamma$)$^{14}$N, while the next step, 
$^{14}$N(p,$\gamma$)$^{15}$O, depletes nitrogen and has a
relatively low cross-section. The final two reactions in the
cycle transform $^{15}$O to $^{12}$C. Since the fourth reaction
runs much slower than the others, 
the cycle achieves equilibrium only when $^{14}$N accumulates to high
levels, and so one effect of the CN
cycle is to convert $^{12}$C to $^{14}$N. 

One issue in nitrogen evolution is to
discover the source of the carbon which is converted into nitrogen, and of
any oxygen which can contribute through the (slow) side chain 
$^{16}$O(p,$\gamma$)$^{17}$F($\beta$$^{+}$,$\nu$)$^{17}$O(p,$\alpha$)$^{14}$N.
For example, stars may produce their own carbon (and 
some oxygen) during helium burning, and the carbon (and perhaps oxygen) is
subsequently processed into $^{14}$N via the CN(O) cycle. 
In this case, nitrogen 
production is independent of the initial composition of the star in which
it is synthesized and is referred to as {\it primary} nitrogen. 
On the other hand, 
stars beyond the
first generation in a galactic system already contain some carbon and oxygen,
inherited from the interstellar medium out of which they formed. The amount
of nitrogen formed from CNO cycling of this material will then 
be proportional 
to its C abundance (and also its O abundance, if the CNO cycling proceeds long
enough to deplete the oxygen) and is known as {\it secondary} nitrogen. 
To zeroth order, then, primary nitrogen production is independent 
of metallicity, 
while secondary production is linearly proportional to metallicity. 
The effects of metallicity on nitrogen production is one facet of nitrogen 
evolution we would like to understand clearly.

A second issue of concern in the origin of nitrogen is 
to identify the portion 
of the stellar mass spectrum which is most responsible 
for $^{14}$N production.
While only massive stars (M$>$8M$_{\sun}$) can produce the necessary internal 
temperatures required to synthesize most heavy elements, 
in the case of nitrogen 
intermediate-mass stars (IMS; 1-8M$_{\sun}$) are also 
massive enough to attain 
required temperatures for its synthesis. In fact, both 
observational evidence and 
theoretical predictions strongly indicate that significant 
amounts of nitrogen are 
produced in intermediate-mass stars (M$<$8M$_{\sun}$), while some is also 
synthesized in massive stars (Maeder 1992; Woosley \& Weaver 1995;
van~den~Hoek \& Groenewegen 1997; 
Henry, Kwitter, \& Bates 2000; 
Henry, Edmunds, \& K{\"o}ppen 2000, hereafter HEK00; 
Marigo 2001; Meynet \& Maeder 2002; 
Siess, Livio, \& Lattanzio 2002).
And since the timescale for IMS evolution is longer 
than for massive stars, nitrogen 
production and ejection into the interstellar medium 
may be significantly delayed 
with respect to oxygen and other heavy 
elements that are produced primarily in 
massive stars. This in turn means that nitrogen evolution 
is affected by the form 
of the initial mass function (IMF), time dependence 
of the star formation rate, 
and the effective stellar nitrogen yields of both massive and 
intermediate-mass stars. 

Local measurements of emission line regions have examined the abundance
of nitrogen relative to O and other $\alpha$-elements as a function
of metallicity (see Henry \& Worthey 2000 for a compilation).
These observations have provided evidence for the contributions of both the 
primary and secondary nitrogen mechanisms 
described above to the cosmic abundance evolution of this element. 
For example, one observes (i) a 'plateau' of \nalph\ measurements
at [\alphH]~$<-1$ consistent with primary nitrogen 
formation, i.e., independent
of metallicity; and 
(ii) a rise in \nalph\ with increasing metallicity above [\alphH]~$\approx -1$
suggestive of secondary nitrogen formation.
Similar observations can be performed with the damped \lya systems
(Pettini et al.\ 1995; Lu et al.\ 1998, hereafter L98; Centuri\'on et al.\ 1998).
Because damped systems tend to have low metallicity ($Z < Z_\odot/10$)
their \nalph\ measurements probe the primary regime of nitrogen
production. In addition, since these systems are relatively young and 
may not have achieved a stage of steady evolution because of differences in 
stellar evolution time scales over the stellar mass spectrum, 
N/$\alpha$ may allow us to gauge the effects of time 
delay and the role of IMS 
in nitrogen production.  And unlike the $\alpha$/Fe ratios described above, 
the \nalph\ observations are largely free 
of the uncertainties due to differential
depletion because of the mild refractory nature of N, O, S, and Si.
In fact the most significant source of uncertainty is due to the effects of
photoionization: measurements of ions \nti, \oi, \suii, H$^0$, and
\siii\ must be converted to elemental abundances.

The initial studies of \nalph\ in damped systems suggested a significant
scatter in the \nalph\ ratios (L98) with a few systems showing values below
the plateau defined by low metallicity H~II regions.
These observations are difficult to interpret as either purely 
primary or secondary nitrogen formation and a combination of the two
channels was suggested (L98).
In this paper, we will reexamine these conclusions.
We present \nnew\ nitrogen measurements from our UCSD HIRES/Keck~I
damped \lya database (Prochaska et al.\ 2001; hereafter, P01).
and combine these data with \nstud\ from 
previous studies and an additional \nlit\ measurements from the recent
literature.  We restrict our discussion to high resolution
echelle observations because (1) the N\,I transitions lie within the \lya
forest where line-blending is important; 
(2) several N\,I transitions are subject to line saturation;
and (3) the N\,I 1134 and 1200 triplets are very closely spaced.
We limit the analysis to the damped \lya systems where the effects of
photoionization are more likely to be small.
We synthesize the abundance measurements from these various
studies, investigate the effects of photoionization, and present a
homogeneous, uniform sample of \nalph\ and \alphH\ measurements.  We then
consider various models of N production and describe the implications
for the star formation histories within the damped \lya systems.

\section{THE UCSD HIRES DATABASE}
\label{sec:data}

All of the systems comprising the UCSD database were observed on the
Keck~I telescope with the HIRES spectrograph \citep{vogt94}.  Most of the
observations are described in P01 and we refer the reader
to that paper for more details on the acquisition and 
processing of the data.  For several of the systems presented in this
section, we also rely upon measurements made with the UVES spectrograph
\citep{dekker00} on the VLT.
In the following subsections, we present the
nitrogen transitions as well as other transitions relevant to our
analysis. These include intermediate-ion transitions
like Fe\,III~1122 which help reveal the ionization state of the gas and
$\alpha$-element transitions like Si\,II~1808 which are essential for
assessing the enrichment history.  The following sub-sections present
the figures and ionic column density measurements for each system.
For ions with multiple transitions measured, 
we report the variance-weighted mean.
Table~\ref{tab:fosc} lists the atomic data adopted throughout the paper where
column~2 lists the rest wavelength, column~3 gives the oscillator strength,
and column~4 lists the reference.
Complete abundance measurements made from the new observations will be
provided in our abundance 
database\footnote{http://kingpin.ucsd.edu/$\sim$hiresdla}.

\begin{table}[ht]\footnotesize
\begin{center}
\caption{{\sc ATOMIC DATA \label{tab:fosc}}}
\begin{tabular}{lcccc}
\tableline
\tableline
Transition &$\lambda$ &$f$ & Ref\\
\tableline
   OI 976  &  976.4481 & 0.00330000 &  1  \\
 NIII 989  &  989.7990 & 0.10660000 &  1  \\
 SiII 989  &  989.8731 & 0.13300000 &  1  \\
 SIII 1012 & 1012.5020 & 0.03550000 &  1  \\
   OI 1039 & 1039.2304 & 0.00919700 &  1  \\
  ArI 1048 & 1048.2199 & 0.26280000 &  3  \\
 FeII 1063 & 1063.1760 & 0.06000000 &  1  \\
 FeII 1064 & 1063.9718 & 0.00371800 &  2  \\
  ArI 1066 & 1066.6600 & 0.06747000 &  3  \\
  NII 1083 & 1083.9900 & 0.10310000 &  1  \\
 FeII 1121 & 1121.9748 & 0.02020000 &  2  \\
FeIII 1122 & 1122.5260 & 0.16200000 &  3  \\
   NI 1134 & 1134.1653 & 0.01342000 &  1  \\
   NI 1134 & 1134.4149 & 0.02683000 &  1  \\
   NI 1134 & 1134.9803 & 0.04023000 &  1  \\
 FeII 1143 & 1143.2260 & 0.01770000 &  2  \\
 FeII 1144 & 1144.9379 & 0.10600000 &  2  \\
 SIII 1190 & 1190.2080 & 0.02217000 &  1  \\
 SiII 1190 & 1190.4158 & 0.25020000 &  1  \\
 SiII 1193 & 1193.2897 & 0.49910000 &  1  \\
   NI 1199 & 1199.5496 & 0.13000000 &  1  \\
   NI 1200 & 1200.2233 & 0.08620000 &  1  \\
   NI 1200 & 1200.7098 & 0.04300000 &  1  \\
SiIII 1206 & 1206.5000 & 1.66000000 &  1  \\
 HI-A 1215 & 1215.6701 & 0.41640000 &  1  \\
  SII 1250 & 1250.5840 & 0.00545300 &  1  \\
  SII 1253 & 1253.8110 & 0.01088000 &  1  \\
  SII 1259 & 1259.5190 & 0.01624000 &  1  \\
   OI 1302 & 1302.1685 & 0.04887000 &  1  \\
 SiII 1304 & 1304.3702 & 0.09400000 &  4  \\
 AlII 1670 & 1670.7874 & 1.88000000 &  1  \\
 SiII 1808 & 1808.0130 & 0.00218600 & 11  \\
AlIII 1854 & 1854.7164 & 0.53900000 &  1  \\
AlIII 1862 & 1862.7895 & 0.26800000 &  1  \\
\tableline
\end{tabular}
\tablerefs{Key to References -- 1:
\cite{morton91}; \\
2: \cite{howk00}; 3:
\cite{morton01}; \\
4: \cite{tripp96}; 11: \cite{bergs93}}
\end{center}
\end{table}

With few exceptions, we calculate ionic column densities with the
same procedures as described in P01, i.e., we implement the apparent
optical depth method \citep{sav91} for the majority of measurements.
This technique, however, cannot be applied to transitions where line-blending 
or saturation is significant.  This occurs more
frequently for transitions presented in this paper because the majority
arise within the \lya forest.  For these cases, we either present 
upper limits on the measured column densities or have performed a 
full line-profile analysis with the VPFIT software package kindly
provided by R. Carswell and J. Webb.  Note that the tables report
[X/H] measurements for the low-ion species having adopted the 
meteoritic solar abundances from \cite{grvss96} 
except for N and Ar where we adopt their photometric values and O
where we adopt the revised value from \cite{holweger01}.

\subsection{PH957, $z$ = 2.309}

Although this system is included in the UCSD/HIRES database, our 
observations do not cover the N\,I transitions.  Therefore, we
adopt the measurements of \nti\ and several other ions 
from the UVES observations studied in \cite{dessauges02b}: 
$\log \N{N^0} = 15.03 \pm 0.02$, $\log \N{HI} = 21.37 \pm 0.08$, 
$\log \N{S^+} = 15.11 \pm 0.01$.  

\begin{figure*}
\begin{center}
\includegraphics[height=8.5in, width=6.0in]{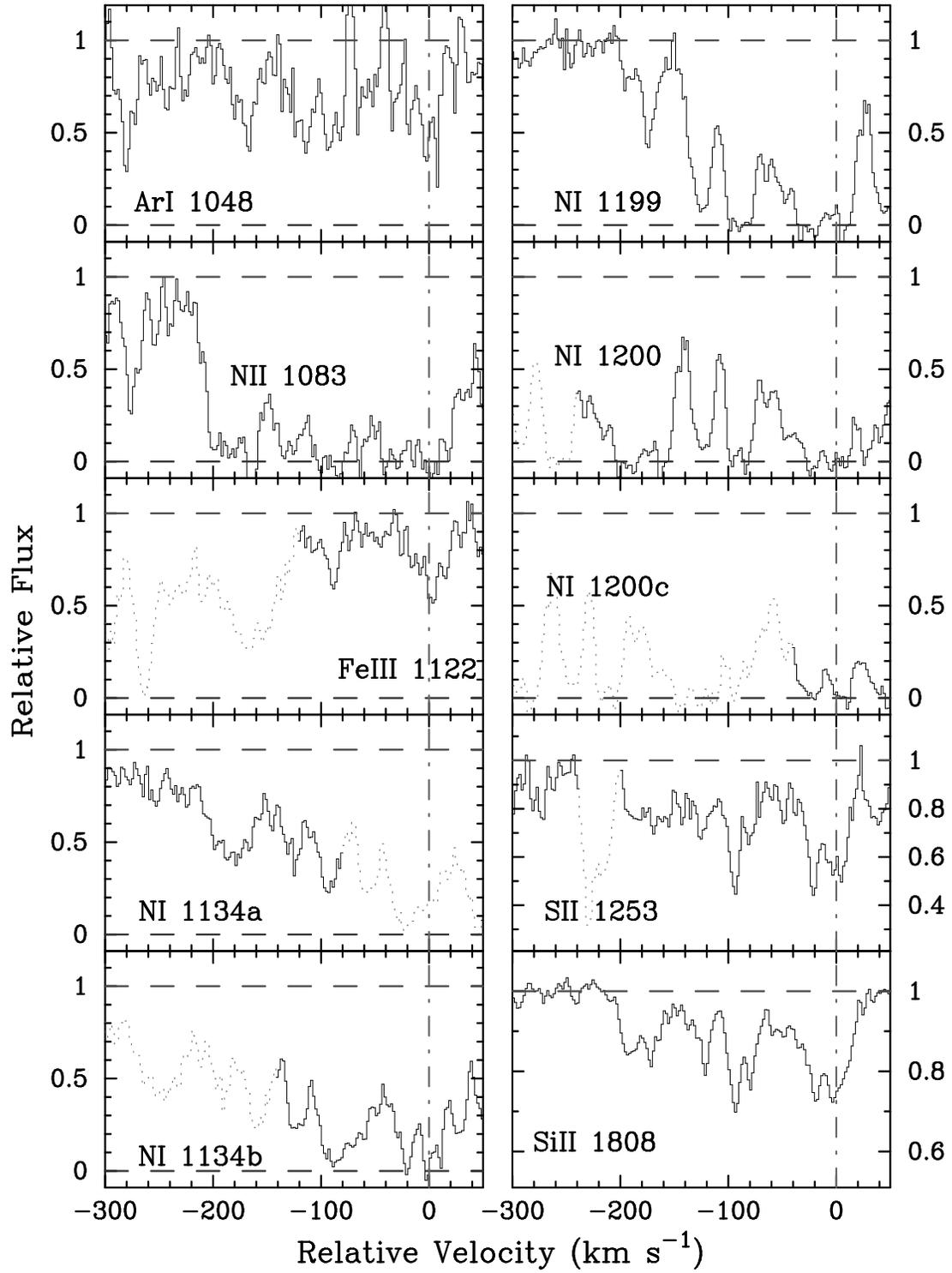}
\caption{Velocity plot of N\,I and intermediate-ion transitions for the 
damped \lya system at $z = 2.463$ toward Q0201$+$36.  
For comparison, we also plot the Si~II 1808 profile.
The vertical line at $v=0$ corresponds to $z = 2.4628$.}
\label{fig:0201+36}
\end{center}
\end{figure*}

\begin{table}[ht]\footnotesize
\begin{center}
\caption{ {\sc
IONIC COLUMN DENSITIES: Q0201+36, $z = 2.463$ \label{tab:Q0201+36_2.463}}}
\begin{tabular}{lcccc}
\tableline
\tableline
Ion & $\lambda$ & AODM & $N_{\rm adopt}$ & [X/H] \\
\tableline
HI &1215.7 & $20.380  \pm 0.045  $ \\
N  I  &1134.1&$$&$>15.000$&$>-1.350$\\  
N  II &1083.9&$>15.303$\\  
Si II&&&$15.532 \pm  0.010$&$-0.408 \pm  0.046$\\  
S  II &1253.8&$15.287 \pm  0.011$&$15.287 \pm  0.011$&$-0.293 \pm  0.046$\\  
Ar I  &1048.2&$14.077 \pm  0.030$&$14.077 \pm  0.030$&$-0.823 \pm  0.054$\\  
Fe II&&&$15.008 \pm  0.004$&$-0.872 \pm  0.045$\\  
Fe III&1122.5&$13.855 \pm  0.023$\\  
\tableline
\end{tabular}
\end{center}
\end{table}

\subsection{Q0201$+$36, $z$ = 2.463}

We present a series of metal-line profiles for this damped system
in Figure~\ref{fig:0201+36} including N\,I, Fe\,III, and Ar\,I.
Because the velocity profiles extend over 200 km/s, the two N\,I
triplets are significantly self-blended.  To estimate the
\nti\ column density, we have performed a line-profile analysis of these
transitions with the VPFIT software package.  
Unfortunately, the blending is so severe and complicated that we could
not achieve a satisfactory solution.  In turn, we can only place a very
conservative lower limit of $\N{N^0} > 10^{15} \cm{-2}$.
Note, the $\N{Fe^{++}}$ value listed in Table~\ref{tab:Q0201+36_2.463} refers
to the velocity region $-120 \mkms < v < 40 \mkms$ which does not include
the strong feature at $v \approx -150 \mkms$, a probable 
line-blend.  Similarly, we report an $\N{Fe^+}$ value corresponding to the
same velocity region to enable an evaluation of the \feiii/\feii\ ratio
($\S$~\ref{sec:photo}).

\begin{figure*}
\begin{center}
\includegraphics[height=8.5in, width=6.0in]{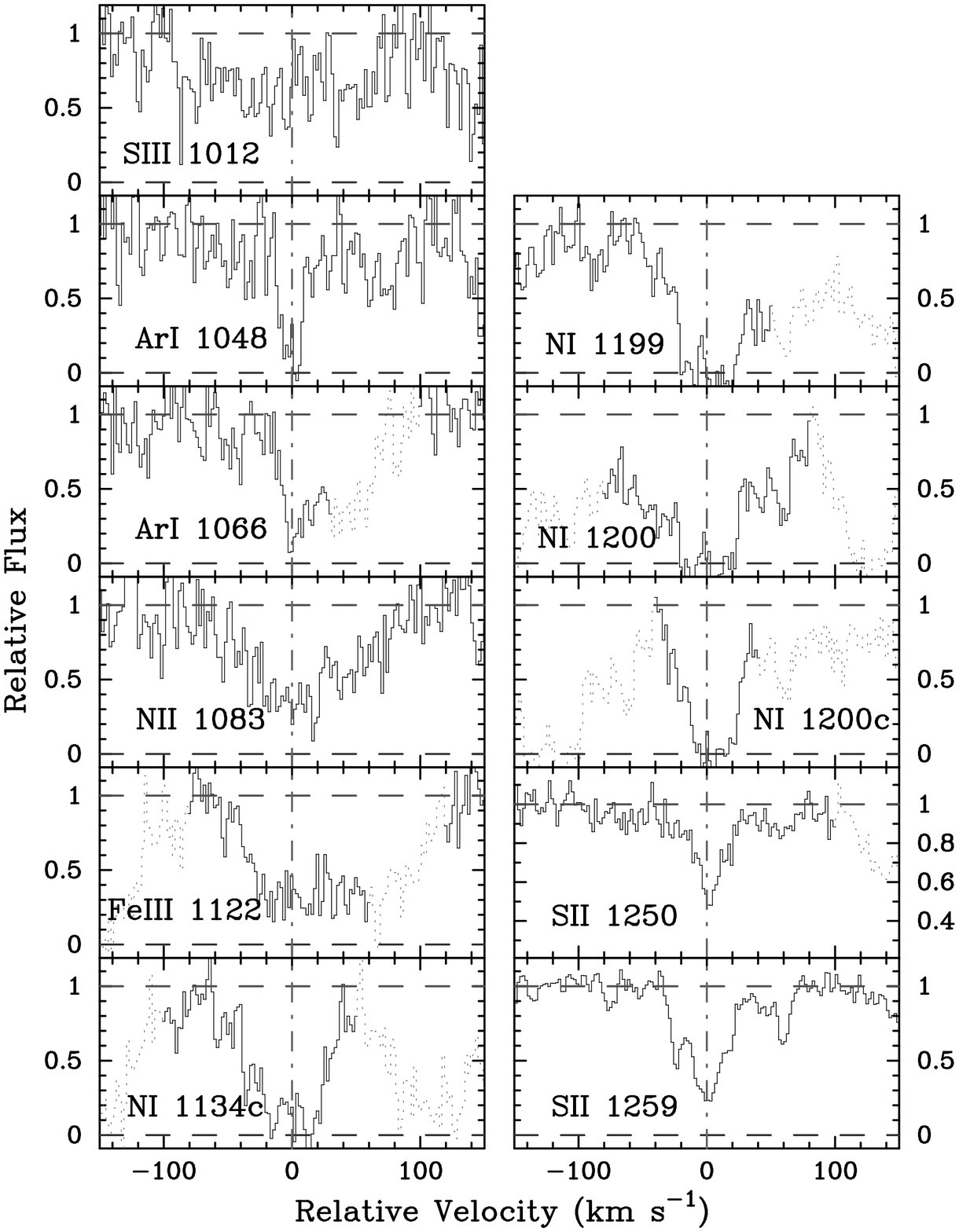}
\caption{Velocity plot of the metal-line transitions for the 
damped \lya system at $z = 3.062$ toward Q0336--01. 
The vertical line at $v=0$ corresponds to $z = 3.062078$.}
\label{fig:0336}
\end{center}
\end{figure*}

\begin{table}[ht]\footnotesize
\begin{center}
\caption{ {\sc
IONIC COLUMN DENSITIES: Q0336-01, $z = 3.062$ \label{tab:Q0336-01_3.062}}}
\begin{tabular}{lcccc}
\tableline
\tableline
Ion & $\lambda$ & AODM & $N_{\rm adopt}$ & [X/H] \\
\tableline
HI &1215.7 & $21.200  \pm 0.100  $ \\
N  I  &1134.9&$>15.037$&$>15.037$&$>-2.133$\\  
N  I  &1200.7&$>15.014$\\  
N  II &1083.9&$<14.371$\\  
S  II &1250.5&$15.118 \pm  0.022$&$14.994 \pm  0.011$&$-1.406 \pm  0.101$\\  
S  II &1259.5&$14.970 \pm  0.012$\\  
S  III&1012.5&$<14.650$&$14.994 \pm  0.011$&$-1.406 \pm  0.101$\\  
Ar I  &1048.2&$>13.939$&$>13.939$&$>-1.781$\\  
Fe II&&&$14.905 \pm  0.033$&$-1.795 \pm  0.105$\\  
Fe III&1122.5&$<14.238$\\  
\tableline
\end{tabular}
\end{center}
\end{table}

\begin{figure*}
\begin{center}
\includegraphics[height=8.5in, width=6.0in]{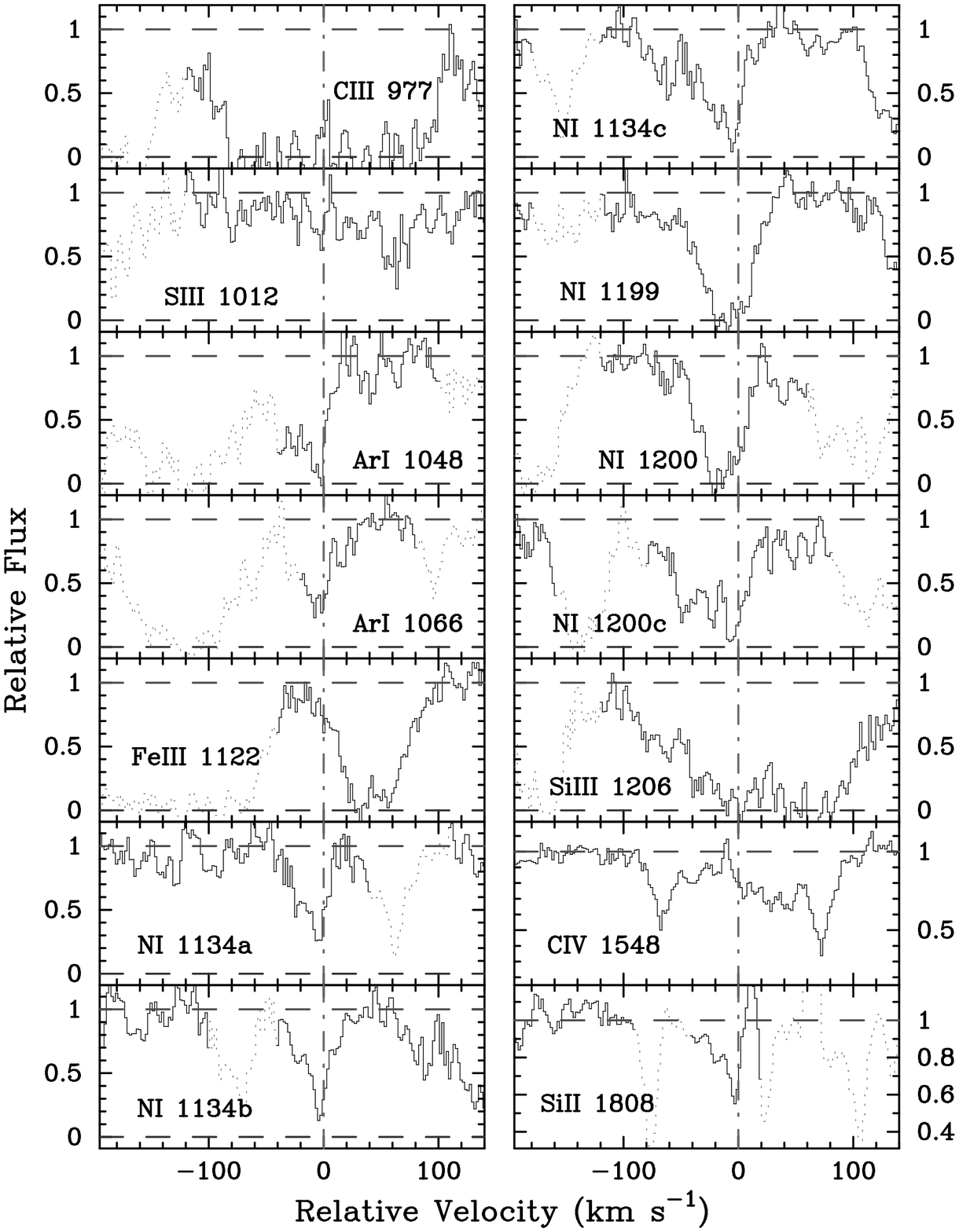}
\caption{Velocity plot of the metal-line transitions for the 
damped \lya system at $z = 3.025$ toward Q0347--38.  
The vertical line at $v=0$ corresponds to $z = 3.0247$.}
\label{fig:0347}
\end{center}
\end{figure*}

\subsection{Q0336$-$01, $z$=3.062}

Figure~\ref{fig:0336} presents the N\,I transitions observed for
the damped \lya system at $z = 3.062$ toward Q0336--01.
Unfortunately all of the N\,I profiles are at least partially
saturated and we can only set a lower limit on the \nti\ column density.  
Table~\ref{tab:Q0336-01_3.062} lists the ionic column density measurements
and the variance-weighted means.
From Fe\,III~1122, we report an upper limit to $\N{Fe^{++}}$ because of the
presence of a significant line blend at $v > 0 \mkms$.
Similarly, we classify our $\N{N^+}$ measurement as an upper limit
because we suspect line-blending with the \lya forest.

\begin{table}[ht]\footnotesize
\begin{center}
\caption{ {\sc
IONIC COLUMN DENSITIES: Q0347-38, $z = 3.025$ \label{tab:Q0347-38_3.025}}}
\begin{tabular}{lcccc}
\tableline
\tableline
Ion & $\lambda$ & AODM & $N_{\rm adopt}$ & [X/H] \\
\tableline
HI &1215.7 & $20.626  \pm 0.005  $ \\
C  III& 977.0&$>14.385$\\  
N  I&&&$14.890 \pm  0.031$&$-1.706 \pm  0.031$\\  
Si II &1808.0&$15.016 \pm  0.026$&$15.016 \pm  0.026$&$-1.170 \pm  0.026$\\  
Si III&1206.5&$>13.447$\\  
S  II&&&$<14.760$&$<-1.066$\\  
S  III&1012.5&$13.849 \pm  0.123$\\  
Ar I  &1048.2&$>14.031$&$14.276 \pm  0.035$&$-0.870 \pm  0.035$\\  
Ar I  &1066.6&$14.276 \pm  0.035$\\  
Fe II&&&$14.503 \pm  0.007$&$-1.623 \pm  0.009$\\  
Fe III&1122.5&$13.140 \pm  0.080$\\  
\tableline
\end{tabular}
\end{center}
\end{table}

\subsection{Q0347$-$38, $z$ = 3.025}
\label{sec:0347ion}

The ionic transitions for the damped system toward Q0347--38
are presented in Figure~\ref{fig:0347}.
This damped system has been carefully analysed by several authors including
\cite{levshakov02} who analysed the high quality UVES/VLT
commissioning data of this quasar. 
In the following, we will adopt their 
HI measurement of $\log \N{HI} = 20.626 \pm 0.005$ and implement
our own metal-line analysis except for nitrogen.
The UVES observations include many more N\,I transitions than our own
and have significantly higher S/N ratio.
We stress that the $\N{Si^+}$ value that we have measured is 
$\approx 0.25$~dex lower than the value listed in \cite{levshakov02}.
We believe the difference is due to a significant telluric line-blend
in the UVES spectrum.  Finally,
the \feiii\ column density reported in Table~\ref{tab:Q0347-38_3.025} refers
to the velocity region $-20 \mkms < v < 20 \mkms$ and may be 
considered an upper limit
because of the likelihood of blending with neighboring, coincident absorption
features.

\begin{figure}[ht]
\begin{center}
\includegraphics[height=5.0in, width=3.5in]{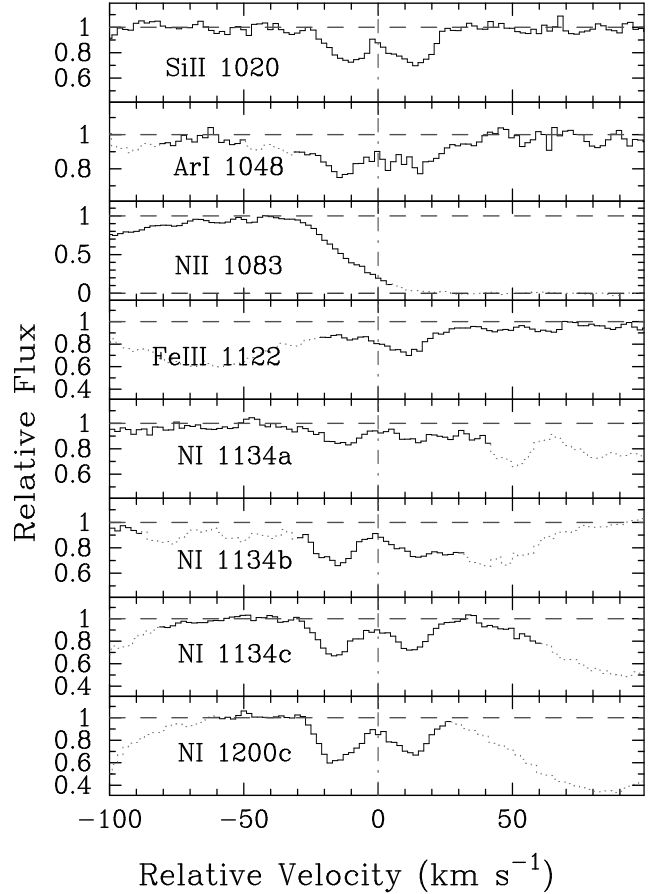}
\caption{Velocity plot of the metal-line transitions for the 
damped \lya system at $z = 3.017$ toward HS0741+47.  
The vertical line at $v=0$ corresponds to $z = 3.017399$.}
\label{fig:0741}
\end{center}
\end{figure}

\begin{table}[ht]\footnotesize
\begin{center}
\caption{ {\sc
IONIC COLUMN DENSITIES: HS0741+4741, $z = 3.017$ \label{tab:HS0741+4741_3.017}}}
\begin{tabular}{lcccc}
\tableline
\tableline
Ion & $\lambda$ & AODM & $N_{\rm adopt}$ & [X/H] \\
\tableline
HI &1215.7 & $20.480  \pm 0.100  $ \\
N  I  &1134.4&$14.158 \pm  0.015$&$13.977 \pm  0.010$&$-2.473 \pm  0.100$\\  
N  I  &1134.9&$13.926 \pm  0.017$\\  
N  I  &1200.7&$14.012 \pm  0.012$\\  
N  II &1083.9&$<14.498$\\  
Si II &1020.6&$14.162 \pm  0.019$&$14.354 \pm  0.003$&$-1.686 \pm  0.100$\\  
Ar I  &1048.2&$13.134 \pm  0.020$&$13.134 \pm  0.020$&$-1.866 \pm  0.102$\\  
Fe II&&&$14.052 \pm  0.005$&$-1.928 \pm  0.100$\\  
Fe III&1122.5&$<13.290$\\  
\tableline
\end{tabular}
\end{center}
\end{table}

\subsection{HS0741$+$47, $z$ = 3.017}

The metal-line 
profiles for this damped system are shown in Figure~\ref{fig:0741} which
includes N\,I, Si\,II, Fe\,III, N\,II, and Ar\,I transitions.  
Because we expect the Fe\,III~1122 and 
N\,II~1083 profiles are dominated by blends with coincident \lya clouds,
we report their column densities as upper limits 
(Table~\ref{tab:HS0741+4741_3.017}).
These limits were derived by integrating the velocity profiles over the
velocity intervals $-20 \mkms < v < 30 \mkms$ and $-40 \mkms < v < 20 \mkms$
respectively.

\begin{figure}[ht]
\begin{center}
\includegraphics[height=4.5in, width=3.5in]{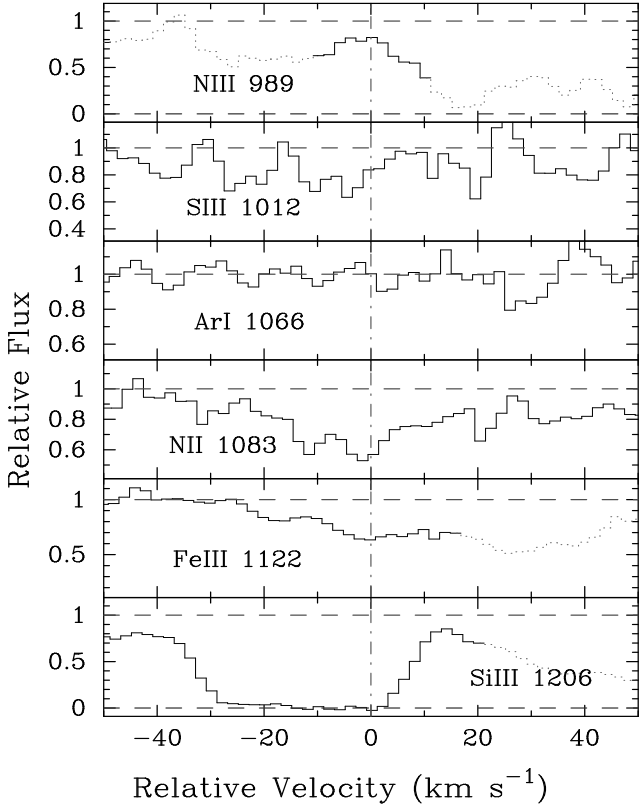}
\caption{Velocity plot of the metal-line transitions for the 
damped \lya system at $z = 3.235$ toward Q0930+28. 
The vertical line at $v=0$ corresponds to $z = 3.2353$.}
\label{fig:0930}
\end{center}
\end{figure}

\begin{table}[ht]\footnotesize
\begin{center}
\caption{ {\sc
IONIC COLUMN DENSITIES: Q0930+28, $z = 3.235$ \label{tab:Q0930+28_3.235}}}
\begin{tabular}{lcccc}
\tableline
\tableline
Ion & $\lambda$ & AODM & $N_{\rm adopt}$ & [X/H] \\
\tableline
HI &1215.7 & $20.300  \pm 0.100  $ \\
N  I  &1134.1&$13.951 \pm  0.057$&$13.740 \pm  0.013$&$-2.530 \pm  0.101$\\  
N  I  &1199.5&$13.659 \pm  0.022$\\  
N  I  &1200.2&$13.762 \pm  0.022$\\  
N  I  &1200.7&$13.884 \pm  0.024$\\  
N  II &1083.9&$13.655 \pm  0.033$\\  
Si II&&&$13.888 \pm  0.021$&$-1.972 \pm  0.102$\\  
Si III&1206.5&$>13.481$\\  
S  III&1012.5&$13.837 \pm  0.089$\\  
Ar I  &1066.6&$<12.960$&$<12.960$&$<-1.860$\\  
Fe II &1121.9&$13.549 \pm  0.081$&$13.695 \pm  0.015$&$-2.105 \pm  0.101$\\  
Fe II &1144.9&$13.767 \pm  0.032$\\  
Fe III&1122.5&$13.141 \pm  0.032$\\  
\tableline
\end{tabular}
\end{center}
\end{table}

\subsection{Q0930$+$28, $z$ = 3.235}

The nitrogen abundance of this damped system was previously analysed
by L98.  In their paper, they report an HI column 
density of $10^{20.18} \cm{-2}$ based on their observations
of the \lya transition.  We find the \lya profile is best fit by
two H\,I absorbers at two redshifts: (1) $z=3.234946$, 
$\N{HI} = 10^{20.3}$, and (2) $z=3.246338$, $\N{HI}=10^{20.2}$.  
We estimate a 0.1~dex error for each of these H\,I measurements.
In the following,
we restrict the analysis to the $z=3.234946$ damped system because
we have significant concerns on the ionization state of the 
'sub-DLA'.  
Figure~\ref{fig:0930} shows the intermediate-ion and Ar\,I transitions 
for this damped system and 
Table~\ref{tab:Q0930+28_3.235} lists the ionic column densities from our
and the L98 analysis. 

\begin{figure}[ht]
\begin{center}
\includegraphics[height=4.5in, width=3.5in]{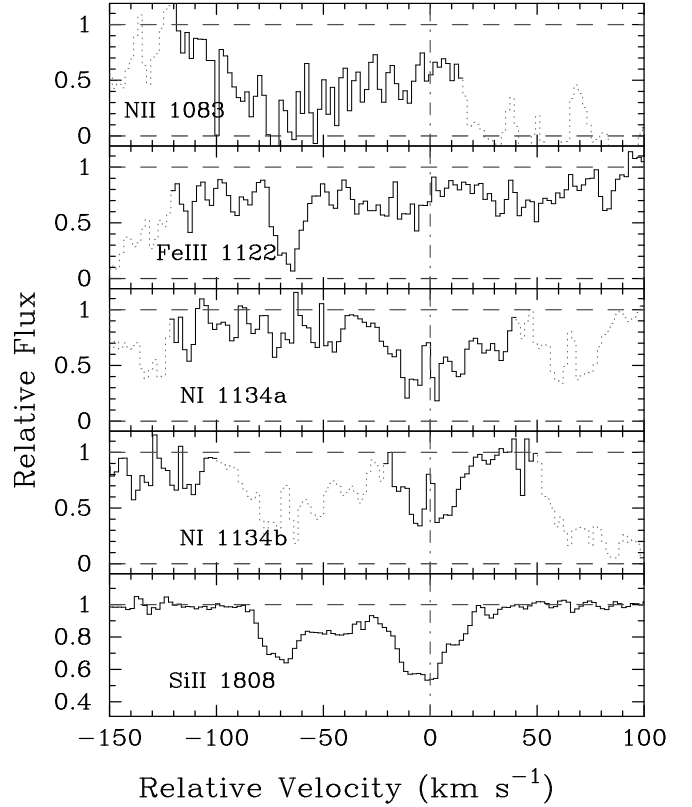}
\caption{Velocity plot of the metal-line transitions for the 
damped \lya system at $z = 2.466$ toward Q1223+17.
The vertical line at $v=0$ corresponds to $z = 2.466083$.}
\label{fig:1223}
\end{center}
\end{figure}

\begin{table}[ht]\footnotesize
\begin{center}
\caption{ {\sc
IONIC COLUMN DENSITIES: Q1223+17, $z = 2.466$ \label{tab:Q1223+17_2.466}}}
\begin{tabular}{lcccc}
\tableline
\tableline
Ion & $\lambda$ & AODM & $N_{\rm adopt}$ & [X/H] \\
\tableline
HI &1215.7 & $21.500  \pm 0.100  $ \\
N  I&&&$14.830 \pm  0.179$&$-2.640 \pm  0.205$\\  
N  II &1083.9&$>14.794$\\  
Si II&&&$15.467 \pm  0.008$&$-1.593 \pm  0.100$\\  
Fe II&&&$15.157 \pm  0.022$&$-1.843 \pm  0.102$\\  
Fe III&1122.5&$<14.082$\\  
\tableline
\end{tabular}
\end{center}
\end{table}

\subsection{Q1223$+$17, $z$ = 2.466}
\label{sec:1223ion}

We reveal the N\,I, Si\,II, and Fe\,III transitions for the damped 
system toward Q1223+17 in Figure~\ref{fig:1223}.  The profiles for
this system are spread over 100 km/s such that the members of the
N\,I~1134 triplet are blended with one another.  We
first estimated the N\,I column density from this triplet by integrating
the apparent optical depth over all 3 transitions and adopting
a summed oscillator strength.
We then performed a detailed line-profile analysis of the lines 
with the VPFIT software package.  Unfortunately, at a S/N ratio of 5 we
could not constrain the $\N{N^0}$ value to better than 0.2~dex.  For
this reason, we have chosen to remove it from our analysis.

\subsection{Q1331$+$17, $z$ = 1.776}

Although we adopt the \siii\ column density presented in P01, we
rely on the analysis of \cite{dessauges02b} for the majority of measurements
considered in this paper.
Their UVES-VLT observations cover the N\,I transitions and 
intermediate-ions which lie blueward of our wavelength coverage. 

\begin{figure*}
\begin{center}
\includegraphics[height=8.5in, width=6.0in]{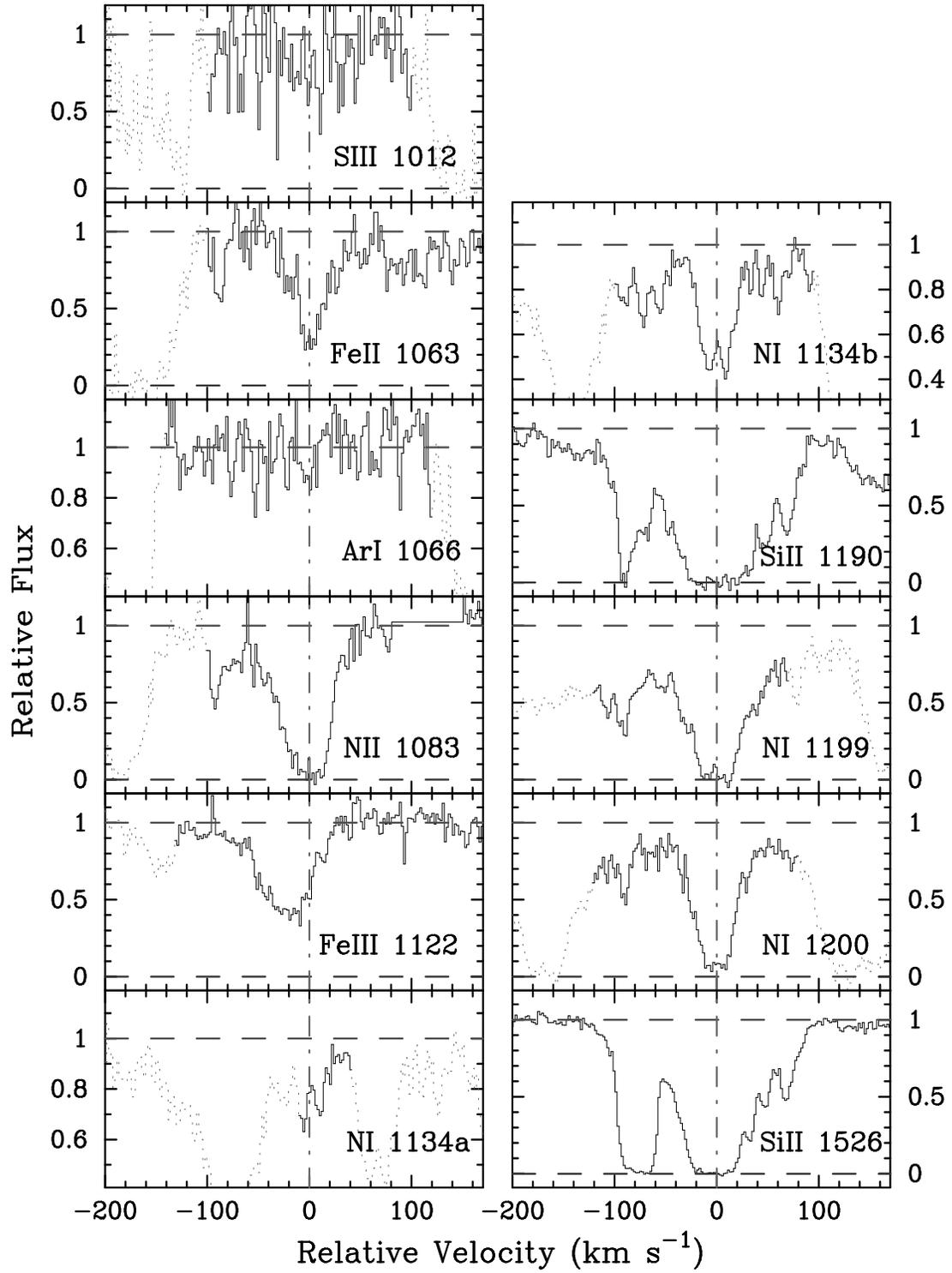}
\caption{Velocity plot of the metal-line transitions for the 
damped \lya system at $z = 2.827$ toward Q1425+60.
The vertical line at $v=0$ corresponds to $z = 2.8268$.}
\label{fig:1425}
\end{center}
\end{figure*}

\begin{table}[ht]\footnotesize
\begin{center}
\caption{ {\sc
IONIC COLUMN DENSITIES: Q1425+6039, $z = 2.827$ \label{tab:Q1425+6039_2.827}}}
\begin{tabular}{lcccc}
\tableline
\tableline
Ion & $\lambda$ & AODM & $N_{\rm adopt}$ & [X/H] \\
\tableline
HI &1215.7 & $20.300  \pm 0.040  $ \\
N  I  &1134.4&$14.707 \pm  0.014$&$14.707 \pm  0.014$&$-1.563 \pm  0.042$\\  
N  I  &1200.2&$>14.623$\\  
N  II &1083.9&$>14.715$\\  
Si II &1190.4&$>14.751$&$>14.826$&$>-1.034$\\  
Si II &1526.7&$>14.826$\\  
S  III&1012.5&$14.157 \pm  0.108$\\  
Ar I  &1066.6&$<13.426$&$<13.426$&$<-1.394$\\  
Fe II &1143.2&$14.255 \pm  0.039$&$14.471 \pm  0.006$&$-1.329 \pm  0.040$\\  
Fe III&1122.5&$<14.018$\\  
\tableline
\end{tabular}
\end{center}
\end{table}
 
\begin{figure*}
\begin{center}
\includegraphics[height=8.5in, width=6.0in]{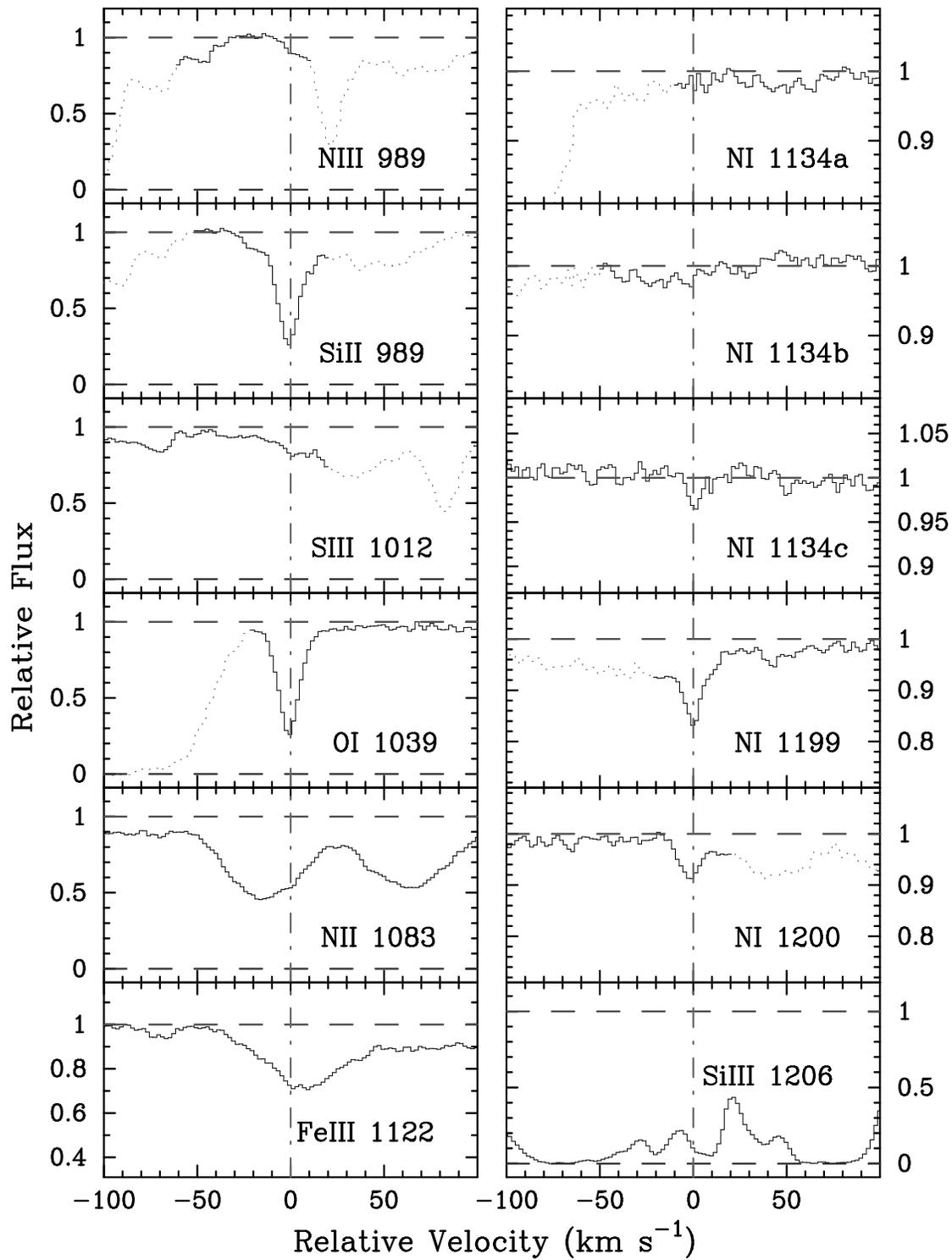}
\caption{Velocity plot of the metal-line transitions for the 
damped \lya system at $z = 2.844$ toward Q1946+76.
The vertical line at $v=0$ corresponds to $z = 2.8443.$}
\label{fig:1946}
\end{center}
\end{figure*}

\subsection{Q1425$+$60, $z$ = 2.827}

L98 presented a lower limit on the \nti\ column density based on their
HIRES spectra of the N\,I~1200 triplet.  
Our independent HIRES observations (Figure~\ref{fig:1425})
provide a measurement of \nti\ which includes the N\,I~1134 triplet
as listed in Table~\ref{tab:Q1425+6039_2.827}.  
Our adopted value is just below their limit and we expect the difference
is due to blending in the \lya forest.
We note that the stronger N\,I transitions exhibit a small but significant
feature at $v \approx -90 \mkms$ which is evident in the Si\,II profiles
as well as the N\,II~1083 profile.  
On the other hand, we suspect that the Fe\,III~1122 profile is significantly
blended with a coincident \lya cloud and place an upper limit to
$\N{Fe^{++}}$.
Unfortunately, both the L98 and our observations only provide a lower limit to
\siii\ which serves as the $\alpha$-element for this system.

\subsection{GB1759$+$75, $z$ = 2.625}

A detailed analysis of this system, including N\,I, was presented
in \cite[][hereafter P02]{pro02}.  
We refer the reader to that paper for figures and tables.

\begin{table}[ht]\footnotesize
\begin{center}
\caption{ {\sc
IONIC COLUMN DENSITIES: Q1946+7658, $z = 2.844$ \label{tab:Q1946+7658_2.844}}}
\begin{tabular}{lcccc}
\tableline
\tableline
Ion & $\lambda$ & AODM & $N_{\rm adopt}$ & [X/H] \\
\tableline
HI &1215.7 & $20.270  \pm 0.060  $ \\
N  I  &1134.1&$<13.214$&$12.588 \pm  0.039$&$-3.652 \pm  0.072$\\  
N  I  &1134.4&$<12.810$\\  
N  I  &1134.9&$12.287 \pm  0.141$\\  
N  I  &1199.5&$<12.860$\\  
N  I  &1200.2&$12.673 \pm  0.038$\\  
N  II &1083.9&$13.568 \pm  0.005$\\  
N  III& 989.7&$12.688 \pm  0.063$\\  
O  I&&&$14.820 \pm  0.007$&$-2.320 \pm  0.060$\\  
O  I  &1039.2&$14.813 \pm  0.008$&$14.820 \pm  0.007$&$-2.320 \pm  0.060$\\  
Si II&&&$13.604 \pm  0.005$&$-2.226 \pm  0.060$\\  
Si III&1206.5&$12.860 \pm  0.005$\\  
S  III&1012.5&$<13.476$\\  
Fe II&&&$13.242 \pm  0.009$&$-2.528 \pm  0.061$\\  
Fe III&1122.5&$<13.216$\\  
\tableline
\end{tabular}
\end{center}
\end{table}

\subsection{Q1946+76, $z$ = 2.844}

The nitrogen abundance of this system has been examined in several
papers (Lu et al.\ 1996; L98).  Our independent observations 
\citep{kirkman97} imply a value in good agreement
with the best of the previous measurements: $\log \N{N^0} = 12.58 \pm 0.04$.
In Figure~\ref{fig:1946}, we present the N\,I transitions as well as
several other metal-lines relevant to the system's ionization state and  
Table~\ref{tab:Q1946+7658_2.844} summarizes the column density measurements.
We report a value for $\N{N^+}$ based on integrating the blended N\,II~1083
profile over the velocity region $-10 \mkms < v < 10 \mkms$.
Finally, the reported \siiii\ value was derived by integrating the Si\,III~1206
profile over the velocity interval $-10 \mkms < v < 10 \mkms$
and might be considered an upper limit given the likelihood of 
line-blending.

\begin{figure}[ht]
\begin{center}
\includegraphics[height=4.5in, width=3.5in]{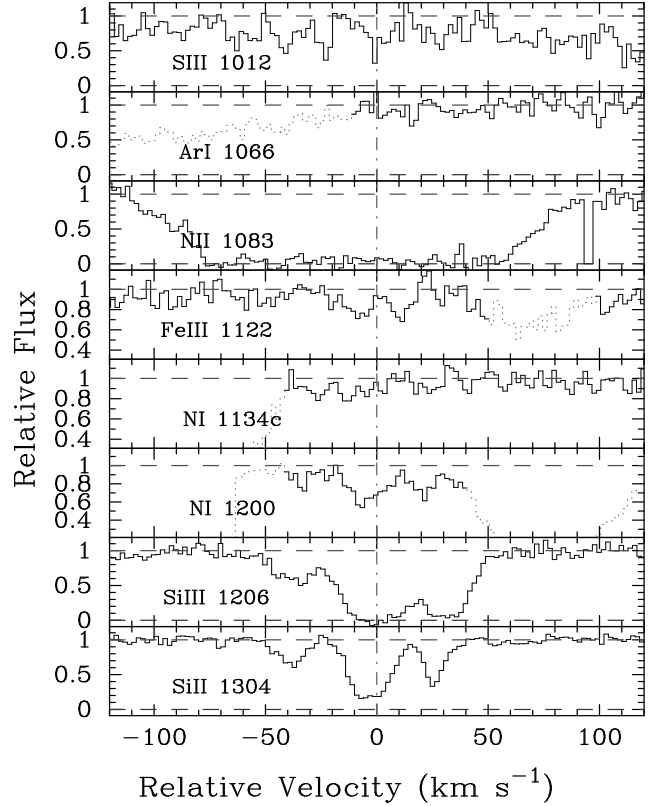}
\caption{Velocity plot of the metal-line transitions for the 
damped \lya system at $z = 2.538$ toward Q2344+12.
The vertical line at $v=0$ corresponds to $z = 2.53790$.}
\label{fig:2344}
\end{center}
\end{figure}

\begin{table}[ht]\footnotesize
\begin{center}
\caption{ {\sc
IONIC COLUMN DENSITIES: Q2344+12, $z = 2.538$ \label{tab:Q2344+12_2.538}}}
\begin{tabular}{lcccc}
\tableline
\tableline
Ion & $\lambda$ & AODM & $N_{\rm adopt}$ & [X/H] \\
\tableline
HI &1215.7 & $20.360  \pm 0.100  $ \\
N  I  &1134.9&$13.742 \pm  0.088$&$13.779 \pm  0.029$&$-2.551 \pm  0.104$\\  
N  I  &1200.2&$13.785 \pm  0.031$\\  
N  II &1083.9&$>15.234$\\  
Si II&&&$14.179 \pm  0.012$&$-1.741 \pm  0.101$\\  
Si III&1206.5&$>13.545$\\  
S  III&1012.5&$14.562 \pm  0.043$\\  
S  III&1190.2&$13.588 \pm  0.089$\\  
Ar I  &1066.6&$<13.256$&$<13.256$&$<-1.624$\\  
Fe II&&&$14.030 \pm  0.032$&$-1.830 \pm  0.105$\\  
Fe III&1122.5&$<13.229$\\  
\tableline
\end{tabular}
\end{center}
\end{table}

\begin{figure*}
\begin{center}
\includegraphics[height=8.5in, width=6.0in]{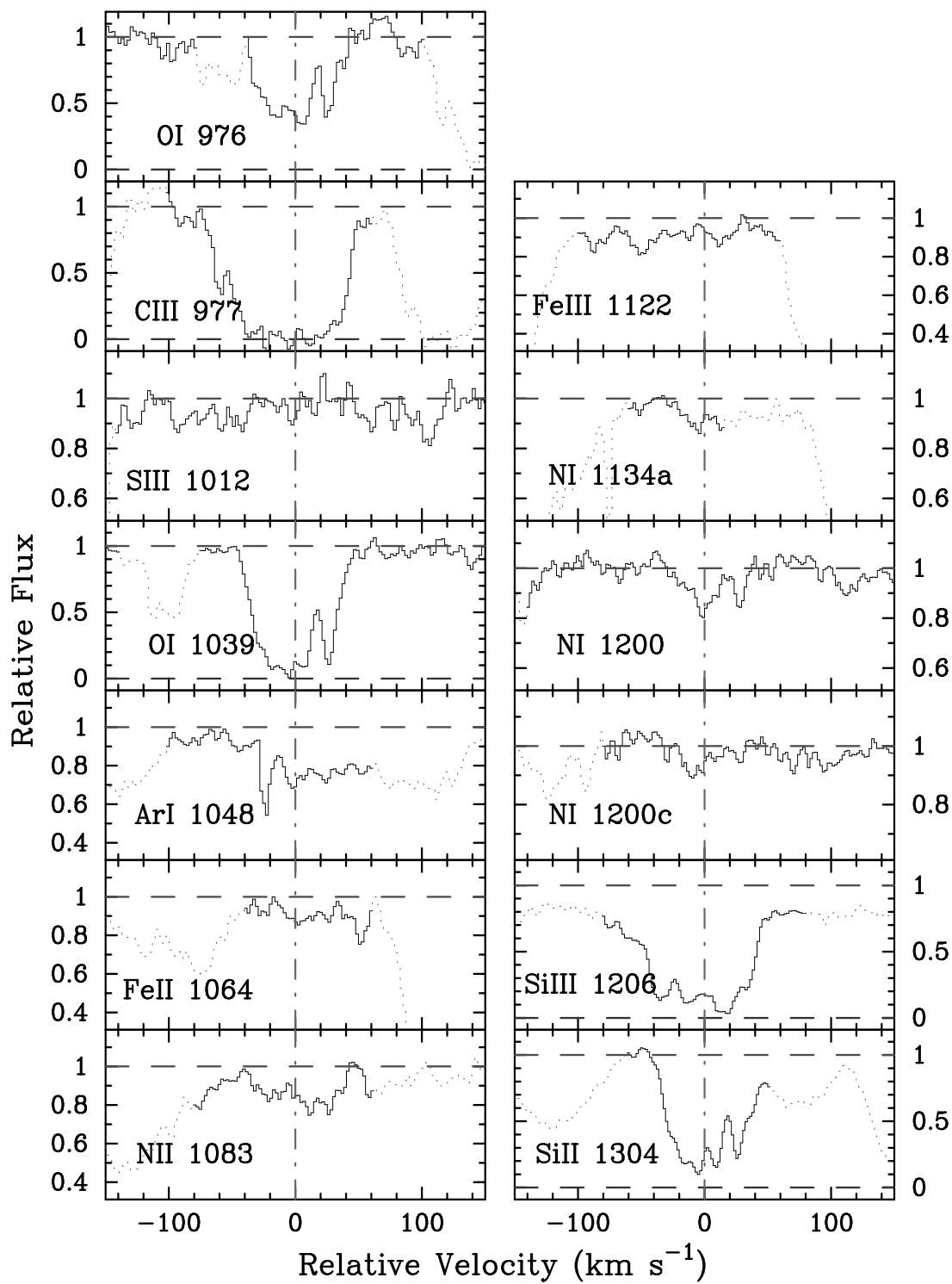}
\caption{Velocity plot of the metal-line transitions for the 
damped \lya system at $z = 3$ toward QXO0001.}
\label{fig:QXO0001}
\end{center}
\end{figure*}

\subsection{Q2344+12, $z$ = 2.538}

L98 presented a measurement of \nti\ based on their HIRES observations.
Our data covers the same transitions as well as several additional
profiles (Figure~\ref{fig:2344}).
Unfortunately, the N\,I~2000a transition is crowded between two \lya 
clouds and it is difficult to accurately assess the \nti\ column density.  
Our adopted value,  $\N{N^0} = 10^{13.779 \pm 0.03} \cm{-2}$, 
is $\approx 0.2$~dex
larger than the value reported by L98.
This difference is substantial and is probably related to measurements of
the continuum level near the N\,I~1200a profile.  Unlike L98, our
observations include the N\,I~1334 triplet and we find an \nti\ column
density from these lines in good agreement with the N\,I~1200a transition.
Table~\ref{tab:Q2344+12_2.538} lists all of the column density measurements
related to our \nalph\ analysis.
This includes a reported upper limit to $\N{Fe^{++}}$ where the
observed profile is most likely blended with several 
coincident absorption lines.
This conclusion stems from the observed differences between the Fe\,III~1122
profile and the low-ion transitions, particularly at $v \approx +5 \mkms$.

\begin{table}[ht]\footnotesize
\begin{center}
\caption{ {\sc
IONIC COLUMN DENSITIES: Q2348-14, $z = 2.279$ \label{tab:Q2348-14_2.279}}}
\begin{tabular}{lcccc}
\tableline
\tableline
Ion & $\lambda$ & AODM & $N_{\rm adopt}$ & [X/H] \\
\tableline
HI &1215.7 & $20.560  \pm 0.075  $ \\
N  I  &1200.2&$<13.235$&$<13.235$&$<-3.295$\\  
Al II&&&$12.657 \pm  0.006$&$-2.393 \pm  0.075$\\  
Al III&1854.7&$12.726 \pm  0.012$\\  
Al III&1862.7&$12.613 \pm  0.027$\\  
Si II&&&$14.203 \pm  0.020$&$-1.917 \pm  0.078$\\  
Si III&1206.5&$>13.499$\\  
S  II&&&$13.725 \pm  0.119$&$-2.035 \pm  0.141$\\  
S  III&1190.2&$<13.945$\\  
\tableline
\end{tabular}
\end{center}
\end{table}

\subsection{Q2348$-$14, $z$ = 2.279}

Most of the important transitions for this damped \lya system were presented
in \cite{pw99} including the N\,I~1200a transition on which our 
$\N{N^0}$ limit is based.  Table~\ref{tab:Q2348-14_2.279} lists the column
densities taken from that paper with revised oscillator strengths where
applicable.  We also report an upper limit to $\N{S^{++}}$ based on the
partially blended S\,III~1190 profile.

\begin{table}[ht]\footnotesize
\begin{center}
\caption{ {\sc
IONIC COLUMN DENSITIES: QXO0001, $z = 3$ \label{tab:QXO0001}}}
\begin{tabular}{lcccc}
\tableline
\tableline
Ion & $\lambda$ & AODM & $N_{\rm adopt}$ & [X/H] \\
\tableline
HI &1215.7 & $20.700  \pm 0.050  $ \\
C  III& 977.0&$>14.171$\\  
N  I  &1134.1&$<14.127$&$13.315 \pm  0.040$&$-3.355 \pm  0.064$\\  
N  I  &1200.2&$13.325 \pm  0.043$\\  
N  I  &1200.7&$13.272 \pm  0.095$\\  
N  II &1083.9&$<13.637$\\  
N  III& 989.7&$>14.738$\\  
O  I  & 976.4&$15.766 \pm  0.024$&$15.766 \pm  0.024$&$-1.804 \pm  0.055$\\  
O  I  &1039.2&$>15.745$\\  
Si II & 989.8&$>14.685$&$14.452 \pm  0.006$&$-1.808 \pm  0.050$\\  
Si II &1193.2&$>14.280$\\  
Si II &1304.3&$14.453 \pm  0.006$\\  
Si III&1206.5&$13.47 \pm 0.10$ & $13.47 \pm 0.10$\\  
S  III&1012.5&$<13.460$\\  
Ar I  &1048.2&$<13.384$&$<13.384$&$<-1.836$\\  
Fe II &1063.9&$<15.092$&$<15.092$&$<-1.108$\\  
Fe III&1122.5&$<12.965$\\  
\tableline
\end{tabular}
\end{center}
\end{table}
 
\subsection{QXO0001, $z$ = 3}

Figure~\ref{fig:QXO0001} presents metal-line profiles for the 
damped \lya system at $z=3$ toward QXO0001 ($\log \N{HI} = 20.7 \pm 0.05$)
which we will discuss further in a future paper.  
Our observations include several intermediate-ion transitions
which we expect are blended with coincident \lya clouds.  
Table~\ref{tab:QXO0001} provides the ionic column density measurements
and limits for all of these transitions.  
Because the Si\,III~1206 transition arises in one wing of 
the damped \lya profile, we first renormalized the continuum surrounding
the Si\,III~1206 profile to account for this depression.
The profile is slightly saturated and one may consider our $\N{Si^{++}}$
value to be formally a lower limit. \\

\vskip 0.4in

\section{DETERMINING [N/$\alpha$] AND [$\alpha$/H]}
\label{sec:photo}

The previous section presented all of the systems with observed N\,I
transitions in our HIRES database.  For the analysis which follows,
we have synthesized these observations with all other high precision
nitrogen measurements for damped \lya systems at $z> 1.5$.
This includes the important sample introduced by L98, the system
toward Q0201+11 by \cite{ellison01}, and 
several new measurements 
\citep{molaro00,levshakov02,dessauges02b,lopez02}
obtained with the UVES-VLT spectrograph \citep{dekker00}.
Table~\ref{tab:sum} summarizes all of these observations. 
Column~6 of the table indicates the expected importance of ionization
corrections as described below and column~7 indicates whether the system
is included in the analysis of $\S$~\ref{sec:analysis}.
In general, those systems where we expect ionization corrections will be 
significant but could not reasonably assess their value were eliminated.

\begin{table*}\footnotesize
\begin{center}
\caption{{\sc SUMMARY\label{tab:sum}}}
\begin{tabular}{lccllccl}
\tableline
\tableline
QSO & $z_{abs}$ & $\N{HI}$ & [$\alpha$/H]\tablenotemark{a}
& [N/$\alpha$] & IC\tablenotemark{c} & Included & Ref \\
\tableline
Q0000-2619  &3.390&21.41&$-1.91\pm 0.08$&$-0.74\pm 0.04$&Y&Y&1,2,3   \\  
PH957       &2.309&21.37&$-1.46^b \pm 0.08$&$-0.85\pm 0.02$&Y&Y&4,5     \\  
Q0201+11    &3.387&21.26&$-1.25^b \pm 0.15$&$-0.65\pm 0.16$&?&Y&6       \\  
Q0201+36    &2.463&20.38&$-0.41\pm 0.05$&$>-0.94$&Y&Y&7,8     \\  
J0307-4945  &4.468&20.67&$-1.55\pm 0.12$&$$&N&N&9       \\  
Q0336-01    &3.062&21.20&$-1.41^b \pm 0.10$&$>-0.73$&Y&Y&8       \\  
Q0347-38    &3.025&20.63&$-1.17\pm 0.03$&$-0.54\pm 0.04$&Y&Y&8, 10   \\  
HS0741+4741 &3.017&20.48&$-1.69\pm 0.10$&$-0.79\pm 0.01$&Y&Y&8       \\  
Q0930+28    &3.235&20.30&$-1.97\pm 0.10$&$-0.56\pm 0.02$&N&Y&8, 11   \\  
Q1055+46    &3.317&20.34&$-1.65\pm 0.15$&$<-0.57$&?&Y&11      \\  
BR1202-07   &4.383&20.60&$-1.81\pm 0.14$&$<-0.46$&?&Y&11      \\  
Q1223+17    &2.466&21.50&$-1.59\pm 0.10$&$-1.05\pm 0.18$&N&N&8       \\  
Q1331+17    &1.776&21.18&$-1.45\pm 0.04$&$-0.49\pm 0.11$&?&Y&5       \\  
Q1425+6039  &2.827&20.30&$>-1.03$&$<-0.53$&N&N&8, 11   \\  
Q1759+75    &2.625&20.76&$-0.89\pm 0.01$&$-0.68\pm 0.02$&N&Y&12      \\  
Q1946+7658  &2.844&20.27&$-2.23\pm 0.06$&$-1.43\pm 0.04$&Y&Y&8, 11   \\  
Q2212-1626  &3.662&20.20&$-1.90\pm 0.08$&$<-0.69$&?&Y&11      \\  
BR2237-0607 &4.080&20.52&$-1.87\pm 0.11$&$<-0.34$&?&Y&11      \\  
HE2243-6031 &2.330&20.67&$-0.87\pm 0.03$&$-0.89\pm 0.02$&?&Y&13      \\  
Q2343+12    &2.431&20.34&$-0.54\pm 0.10$&$-1.10\pm 0.09$&N&N&5       \\  
Q2344+12    &2.538&20.36&$-1.74\pm 0.10$&$-0.81\pm 0.03$&Y&Y&8       \\  
Q2348-14    &2.279&20.56&$-1.92\pm 0.08$&$<-1.38$&N&Y&2,8,14  \\  
QXO0001      &3    &20.70&$-1.81\pm 0.05$&$-1.45\pm 0.04$&Y&Y&8       \\  
\tableline
\end{tabular}
\end{center}
\tablenotetext{a}{Assumes Si unless noted}
\tablenotetext{b}{Sulfur}
\tablenotetext{c}{See below for a detailed discussion of ionization corrections.}
\tablerefs{Key to References -- 1:
\cite{lu96}; 2: \cite{pw99};
3: \cite{molaro00}; 4: \cite{wolfe94}; 
5: \cite{dessauges02b}; 6: \cite{ellison01}
7: \cite{pw96}; 8: This paper;
9: \cite{dessauges01}; 10: \cite{levshakov02};
11: \cite{lu98}; 12: \cite{pro02};
13: \cite{lopez02}; 14: \cite{ptt95}}
\end{table*}


Given the appropriate wavelength coverage,
high resolution observations of the damped systems can
generally provide precise and accurate ionic column densities for both
nitrogen and an $\alpha$-element.
Before converting these gas-phase measurements into elemental abundances
and \nalph\ ratios,
however, one must consider several factors:
(1) identifying the 'best' $\alpha$-element;
(2) dust depletion; and (3) photoionization.
At present, we believe the first two issues have relatively minor
impact on derivations of \nalph\ and \alphH\ in the damped systems.
In previous analyses and this work too, authors have relied on Si
to serve as the $\alpha$-element because it is the most readily observed.
This raises some concern because
Si is expected to be synthesized in less massive progenitors of
Type~II SN than O or Mg \citep{ww95} and may even be 
substantially produced in Type~Ia SN \citep{matteucci01}.
Therefore, reporting [\nalph] as [N/Si] may complicate comparisons
against theoretical models which generally
reflect nucleosynthesis in the most massive stars.
Empirically, however, we and others have found that [O/Si]~$\approx 0$ and
[S/Si]~$\approx 0$ in the few damped systems where Si and O or S have been
determined.  Therefore, we believe the difference between O and Si
leads to an uncertainty in [\nalph] of less than 0.1~dex.  

\begin{figure*}[ht]
\begin{center}
\includegraphics[height=5.5in, width=4.5in]{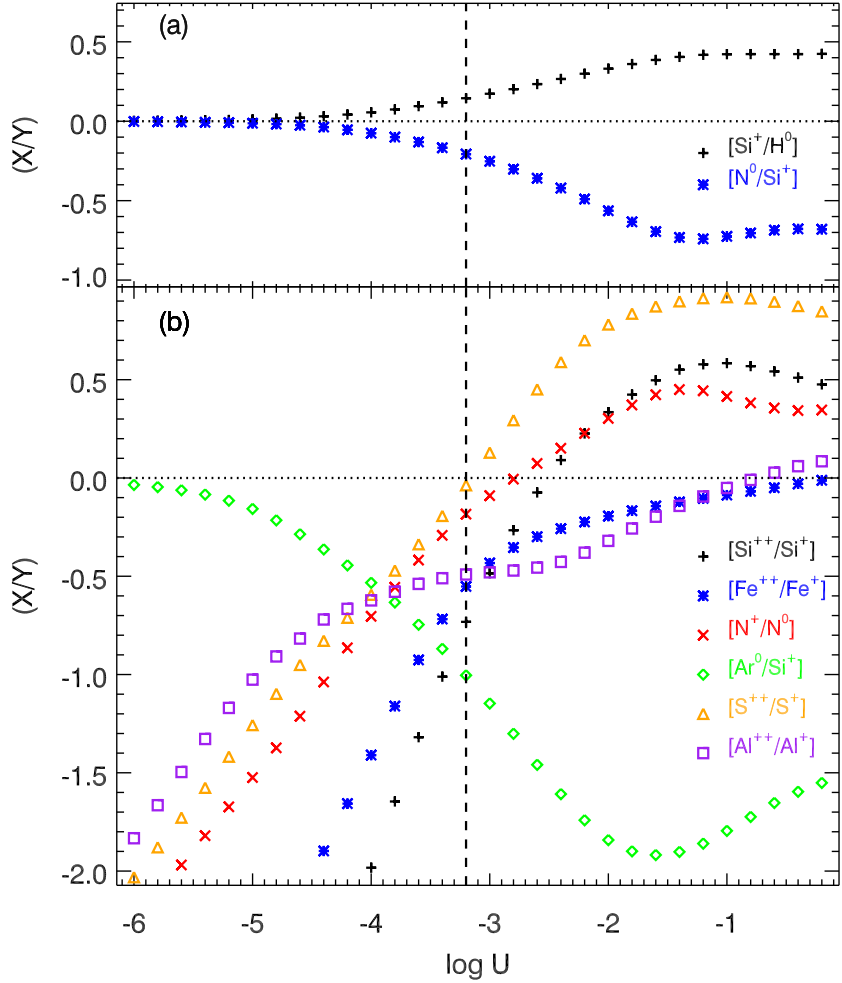}
\caption{Ionic ratios predicted for a plane-parallel slab of
gas with $\N{HI} = 10^{20} \rm{cm^{-2}}$ for a range of ionization
parameter $U$.  These calculations were carried out with the 
CLOUDY software package.  The dotted vertical line indicates the
transition from a predominantly neutral gas where ionization
corrections to [\nti/\siii] are small to a partially ionized
gas where ionization corrections are significant.
Provided the appropriate observations,
the calculations presented in this figure 
can be used to roughly assess
the ionization corrections for a given damped \lya system.
}
\label{fig:cldy}
\end{center}
\end{figure*}

We expect that dust depletion has an even smaller effect on the
[\nalph] and [\alphH] values than uncertainties between the various
$\alpha$-elements.  Although Si is significantly depleted along many
ISM sightlines \citep[e.g.][]{savage96}, the depletion levels 
in the damped \lya systems 
(as inferred from Zn/Fe or S/Fe) are
too small for the differential depletion of Si/N or Si/H
to be important \citep[e.g.][]{vladilo98}.
This is particularly true for the systems presented here where the
Si/Zn and Si/S ratios are very nearly solar (Prochaska \& Wolfe 2002;
but see Ledoux et al.\ 2002 for a counter-example).
The largest corrections one could derive for Si relative to N
and H is +0.2~dex which would
imply slightly lower \nalph\ ratios and a higher \alphH\ value. 

The major sources of systematic error in converting \nti,
\siii, H$^0$, and O$^0$ column densities into elemental abundances
are ionization corrections.  Consider first the ionization balance
of O$^0$, O$^+$, \nti, \ntii, H$^0$, and H$^+$.  Charge-exchange
reactions between (O$^0$, H$^0$) and (\nti, H$^0$) help to regulate
the ionization fractions of these ions such that H$^0$/H $\approx$
\nti/N $\approx$ O$^0$/O when the volume density is large and the
ionizing flux is modest.  If the ratio of ionizing flux to volume
density -- the ionization parameter $U$ -- is large enough, then
this set of equalities breaks down.  Both O$^0$ and \nti\ have
larger cross-sections to photons with $h \nu > 2$~Ryd than \hi\
\citep{sofia98} and these ions will be over-ionized: \nti/N $<$ \hi/H and
\oi/O $<$ \hi/H.  
The likelihood of over-ionization for Ar is even greater as found
in the Galactic ISM \citep{sofia98} and in damped systems (P02).
In contrast, \siii\ can be present in highly ionized regions such
that \siii/Si~$\geq$ \hi/H for nearly all physical conditions.
The net result is that \siii/\hi\ will overestimate Si/H and
\nti/\siii\ will underestimate N/Si {\it if ionization is significant} 
in the damped system, i.e., when $\log U \gg -3$ (see below).
The magnitude of these corrections ranges from a few hundredths dex
in a very neutral gas to several tenths dex in an ionized gas.
To accurately assess these ionization corrections, 
one must examine photoionization diagnostics within each damped system.

Historically, researchers have assumed that ionization corrections
are negligible for the damped systems.  This presumption is physically motivated
by the very large optical depth at energies $h \nu > 1$~Ryd
implied by a sightline with $\N{HI} > 2 \sci{20} \cm{-2}$ \citep{viegas94}.
Detailed theoretical models have drawn similar conclusions 
for the majority of elements observed in
the damped systems \citep{howk99,vladilo01}.  More 
recently, however, \cite{pro02} presented evidence 
for significant photoionization
in $\approx 50\%$ of the gas associated with the damped \lya system at $z=2.62$
toward GB1759+75.  For this gas, the ionization corrections are 
important for ions like \nti\ and \siii\ relative to \hi.  The correction
for \nti/\siii\ in this partially ionized gas is large: 
[\nalph] = [\nti/\siii]~+~0.3~dex.  
Although the authors stress that this particular damped system has
several characteristics which separate it from the majority of damped
systems, its properties highlight the importance of assessing the 
ionization state of each damped system in order to convert
the observed \nti/\siii, \nti/\suii, \oi/\hi, \suii/\hi, and \siii/\hi\ 
ratios to \nalph\ and \alphH.

In Figure~\ref{fig:cldy}, we present a series of ionic ratios 
which qualitatively assess the level of ionization
corrections for gas with a range of ionization states.
These values were calculated with the CLOUDY software package
(v95b; Ferland 2001) assuming a Haardt-Madau extragalactic background
radiation field at $z=2.5$ \citep{haardt96}.
In panel (a) we plot \nti/\siii\ and \siii/H$^0$ versus the ionization
parameter, $U \equiv \phi/(c n_H)$ where $\phi$ is the surface
flux of ionizing photons with $h \nu > 1$~Ryd.  
The vertical dotted line at $\log U = -3.5$ demarcates
the transition from corrections of less than 0.2~dex for [\nti/\siii] to
greater than 0.2~dex.
At this $U$ value, panel (b) shows:
[\feiii/\feii]~$\approx -0.8$,
[\siiii/\siii]~$\approx -1.1$,
[\suiii/\suii]~$\approx -0.35$,
[\ari/\siii]~$\approx -0.7$,
[\aliii/\alii]~$\approx -0.6$,
and
[\ntii/\nti]~$\approx -0.5$.
Systems which exhibit significantly lower values for these ratios
(higher values for \ari/\siii) should be predominantly neutral and require
small corrections for [\nalph] and [\alphH].  Similarly,
systems with larger values of these ratios are presumably partially
ionized and may require significant ionization corrections.
In the following sub-sections, we will describe the ionization 
diagnostics for each of the damped systems in the full nitrogen
sample and evaluate the magnitude of ionization corrections.

\subsection{Q0000$-$26, $z$ = 3.390}

\cite{molaro01} presented \ari\ measurements for this damped \lya system
which indicate a nearly solar Ar/Si ratio.  We agree with their argument
that the gas in this damped \lya system is therefore predominantly neutral.

\subsection{PH957, $z$ = 2.309}

\cite{dessauges02b} will present a detailed analysis of the photoionization
state of this damped system.  Their assessment, based on 
[\ari/\suii]~$\approx -0.2$ and observations of the Fe\,III~1122
transition, is that ionization corrections for \nti\ and \suii\ are small.

\subsection{Q0201$+$11, $z$ = 3.387}

There is no photoionization diagnostic for this damped system, but we
expect that its HI column density is large enough that photoionization
corrections should be small.

\subsection{Q0201$+$36, $z$ = 2.463}

Our new data on this system covers the Fe\,III~1122 transition and provides
a measurement of \ari.  The [\ari/\siii] value of $-0.42 \pm 0.1$~dex 
suggests the possibility of mild photoionization while the observed
\feiii/\feii\ ratio of $-1$~dex indicates such effects
are likely to be small. Altogether we expect ionization corrections 
of 0.1~dex or less and assume no correction.
For this damped system, we measure ions for two $\alpha$-elements, \siii\
and \suii\, whose abundances relative to solar are in reasonably good
agreement.  For [\alphH], we adopt the [Si/H] value because the Si\,II~1808
profile is outside the \lya forest and therefore less likely to be contaminated
by line-blending.

\subsection{J0307--4945, $z$ = 4.466}

\cite{dessauges01} presented a detailed analysis of this damped 
system from their UVES observations which included measurements of
N, Si, O and Fe among other elements.   The authors presented \nti\ column
density measurements for the strongest velocity components
of this kinematically complex system and discussed the relatively low
\nalph\ value.  We are concerned, however, that this gas may
require significant ionization corrections.  We note that for those
components where the O\,I profiles were unsaturated, the measured 
\oi/\siii\ ratios are significantly sub-solar, in some cases by over 1~dex.
This is a clear signature of photoionization in these weaker components
where N\,I is undetected.  Unfortunately, the O\,I profiles are heavily
saturated in those components where N\,I was measured
and it is impossible to accurately assess the photoionization state of
the gas from the data presented in \cite{dessauges01}.
Although we consider it a reasonable possibility that this system does
exhibit a low \nalph\ value (consistent with the low nitrogen sub-sample
discussed in $\S$~\ref{sec:analysis}), we choose to eliminate it from
our sample on the grounds that photoionization may be important.

\subsection{Q0336$-$01, $z$=3.062}

Our observations of the damped \lya system toward Q0336--01 provide a lower
limit on the \ari/\suii\ ratio and place upper limits on the \ntii/\nti\ and 
\feiii/\feii.  The nearly solar [\ari/\suii] ratio and the relatively
small \ntii/\nti\ and \feiii/\feii\ limits all indicate this system is
predominantly neutral.

\subsection{Q0347$-$38, $z$ = 3.025}

To assess the ionization state of this damped
system we focus on the \ari\ measurements made by \cite{levshakov02}.  
Their analysis indicates [\ari/\siii]~$> -0.2$ implying this 
system is neutral.  
This conclusion is supported by our measured ratio of 
\feiii/\feii~$= -1.4$~dex which may be considered an upper limit
($\S$~\ref{sec:0347ion}).
We also report a \suiii/\suii\ value of $\lesssim -0.8$~dex which is
consistent with this conclusion.

It is worth noting, that the C\,III~977,
Fe\,III~1122, S\,III~1012, and Si\,III~1206 profiles all exhibit 
significant absorption at $v > 20 \mkms$.  Although 
this absorption may result from coincident \lya clouds, 
we contend the absorption is more likely due to a significantly ionized
gas associated with the damped system.  
This component does not give rise to significant low-ion absorption,
however, and therefore has no effect on our \nalph\ analysis.

\subsection{HS0741$+$47, $z$ = 3.017}

Our spectrum of HS0741+47 includes coverage of several ionization
diagnostics for this damped \lya system including Ar\,I, Fe\,III, and
N\,II transitions.  Unfortunately, the Fe\,III 1122 and N\,II 1083 transitions
appear to be blended with coincident \lya forest clouds
(Figure~\ref{fig:0741}) and the resulting upper limits are not too meaningful.
On the other hand, [\ari/\siii]~$= -0.18$ and we contend the system is
primarily neutral with ionization corrections for [\nalph]
and [\alphH] less than 0.1~dex.

\subsection{Q0930$+$28, $z$ = 3.235}

The strongest constraints on the ionization state of this system are the
\ntii/\nti\ and \siiii/\siii\ ratios.  The N\,II~1083 profile may be mildly
blended with a coincident \lya forest cloud, but we expect 
\ntii/\nti~$\approx -0.2$~dex which implies a modest ionization correction
for nitrogen.  Similarly, the saturated Si\,III~1206 profile yields a
lower limit to \siiii/\siii\ of $-0.25$~dex consistent with this system
being partially ionized.  Altogether, we
contend the N/$\alpha$ and $\alpha$/H values would require significant
ionization corrections and we do not include this system in our analysis.
In passing, however, we note that the uncorrected [\nalph] and
[\alphH] values are typical for the majority of our sample.

\subsection{Q1055+46, $z$ = 3.317}

We have no ionization diagnostics for this system.

\subsection{BR1202$-07$, $z$ = 4.383}

We have no ionization diagnostics for this system.

\subsection{Q1223$+$17, $z$ = 2.466}
\label{sec:1223pht}

As described in $\S$~\ref{sec:1223ion}, we have been unable to measure
an accurate \nti\ column density from our dataset and have not included
this system.  
In passing, we note the Fe\,III and N\,II profiles indicate the gas is
predominantly neutral at $v \approx 0 \mkms$ but suggest the gas at
$v \approx -70 \mkms$ could be partially ionized.
The latter point hinges on whether the feature observed at $v \approx -70 \mkms$
in the Fe\,III~1122 profile is the result of a blend.

\subsection{Q1331$+$17, $z$ = 1.776}

A complete analysis of this system's ionization state will be
presented in \cite{dessauges02b}.  
Their UVES observations of Fe\,III~1122 indicate the gas giving
rise to the N\,I profiles is predominantly neutral.

\subsection{Q1425$+$60, $z$ = 2.827}

This system exhibits several diagnostics which indicate photoionization
is likely to be important.  In particular, we find [\ari/\siii]~$< -0.5$~dex
and \ntii/\nti\ $> 0$~dex.  Although the N\,II~1083 profile is mildly 
saturated and therefore resembles a typical \lya cloud, we are confident
it is not a coincident absorption line because the profile also exhibits
a velocity component at $v \approx -90 \mkms$ which coincides with a
similar feature in the low-ion profiles.  
Analysing this single velocity component,
we find \ntii/\nti~$\approx 0$~dex consistent with the total $\N{N^+}$
and $\N{N^0}$ values.
Altogether we expect the measured \nti\ column density significantly 
underestimates the nitrogen elemental abundance.
We choose to remove it from the analysis noting, however, that it
does not appear to exhibit a low \nalph\ ratio.
In fact, if ionization corrections to \nti\ and \siii\ are significant,
this system may exhibit a nearly solar \nalph\ ratio.

\subsection{GB1759$+$75, $z$ = 2.625}

A detailed photoionization treatment of this damped system has been
presented by P02.  We concluded that roughly half of 
the gas is partially ionized with \nti/\siii\ requiring significant 
ionization corrections.  The remaining gas, 
meanwhile, is predominantly neutral.
We are able to estimate the ionization corrections to [\nalph] and [\alphH]
for the entire system and have therefore included it within our analysis.
The corrections we adopt are +0.15~dex to [\nalph] and $-0.1$~dex to [\alphH]
as measured from [Si/H].

\subsection{Q1946+76, $z$ = 2.844}

This well studied damped \lya system has a relatively low HI column
density and one of the lowest \nti/\siii\ ratios observed.  These factors
raise concerns that its gas may be primarily ionized and that the low
inferred \nalph\ value has resulted from photoionization.  This
concern is supported by the slightly sub-solar \oi/\siii\ ratio,
but we place a strict upper limit on \siiii/\siii~$< -0.75$~dex which
suggests ionization corrections are not extreme.  
Unfortunately, the N\,II~1083 profile is blended and does not provide
a useful limit on \ntii/\nti.
Nevertheless, the system may require an ionization correction
for [\nalph] (see Figure~\ref{fig:cldy})
and we have remaining concerns on its ionization state.
In the following, however, we proceed without adopting any correction.

\subsection{Q2212$-$16, $z$ = 3.662}

There is no ionization diagnostic for this damped system.

\subsection{BR2237$-$06, $z$ = 4.080}

There is no ionization diagnostic for this damped system.

\subsection{HE2243$-$60, $z$ = 3.149}

This new damped \lya system \citep{lopez02} exhibits competing 
photoionization diagnostics.  Although the system shows a significantly
sub-solar \ari/\siii\ ratio, [\ari/\siii]~$= -0.5$, \cite{lopez02}
examined the \alii\ and \feii\ column densities and argued that 
ionization corrections are likely to be small.  We proceed under the
assumption that ionization corrections to \nalph\ are small.

\subsection{Q2343$+$12, $z$ = 2.431}

\cite{dessauges02b} will present a full analysis on the photoionization
of this system.  We note here that they find a very low \ari/\siii\ ratio
and strong \ntii\ absorption indicating photoionization corrections will
be important.  As such, we have not included this system in our analysis.

\subsection{Q2344+12, $z$ = 2.538}

Although our observations cover several ionization diagnostics, the
only valuable ratio is \feiii/\feii; N\,II~1083 is severely blended
and the \ari\ transitions provide only an upper limit to $\N{Ar^0}$.
The upper limit on \feiii/\feii~$< -0.8$ suggests the system is probably
neutral but the \nti\ and \siii\ column densities
could be subject to small ($\lesssim 0.1$~dex) ionization corrections.
In the following, we adopt no corrections.

\subsection{Q2348$-$14, $z$ = 2.279}

The very low upper limit on \nti/\siii\ for this damped system raises
concerns that photoionization may be important.  Unfortunately, the
existing spectra do not cover the Fe\,III or N\,II profiles.
We do measure \aliii/\alii $= -0.55$~dex but
this diagnostic is relatively insensitive
to the photoionization state (Figure~\ref{fig:cldy}) and does not provide
a particularly meaningful constraint.  
We also place a lower limit on \siiii/\siii\ of $-0.7$~dex which may
be influenced by a line-blend in the \lya forest.
Unfortunately, these diagnostics are inconclusive.
Most worrisome is the fact that the Si\,IV and C\,IV profiles
exhibit significant features at essentially identical velocity as the
low-ion absorption indicating ionized gas is at this velocity.
As discussed in P02 and \cite{wp00a}, 
it is unlikely that the gas responsible for the
high-ion profiles is associated with the low-ion profiles in damped
\lya systems.  
Nevertheless, we have remaining concerns regarding the ionization
state of this system.  On the other hand, the \nti\ column density
is an upper limit and we are reasonably confident that 
the \nalph\ is very low in this damped system.

\subsection{QXO0001, $z$ = 3}

Although our observations provide several ionization diagnostics for
this system, it is difficult to draw a definite conclusion on its
ionization state.  The N\,II~1083 profile is blended 
and places a large upper limit on \ntii/\nti.  The Ar\,I transition
is also blended and provides an upper limit [\ari/\siii]~$< 0$~dex 
which is consistent with both a neutral and ionized gas.  
Meanwhile, the partially saturated
Si\,III~1206 transition implies \siiii/\siii~$\gtrsim -1$ and we
place an upper limit on \feiii/\feii\ of $-1$~dex.  These latter two
diagnostics suggest the system is primarily neutral, but a mild ionization
correction to \nalph\ is allowed (up to 0.2~dex).  We 
proceed assuming a +0.1~dex correction to [\nalph] and no correction
to [\alphH].

\begin{figure*}[ht]
\begin{center}
\includegraphics[height=6.0in, width=4.5in, angle=-90]{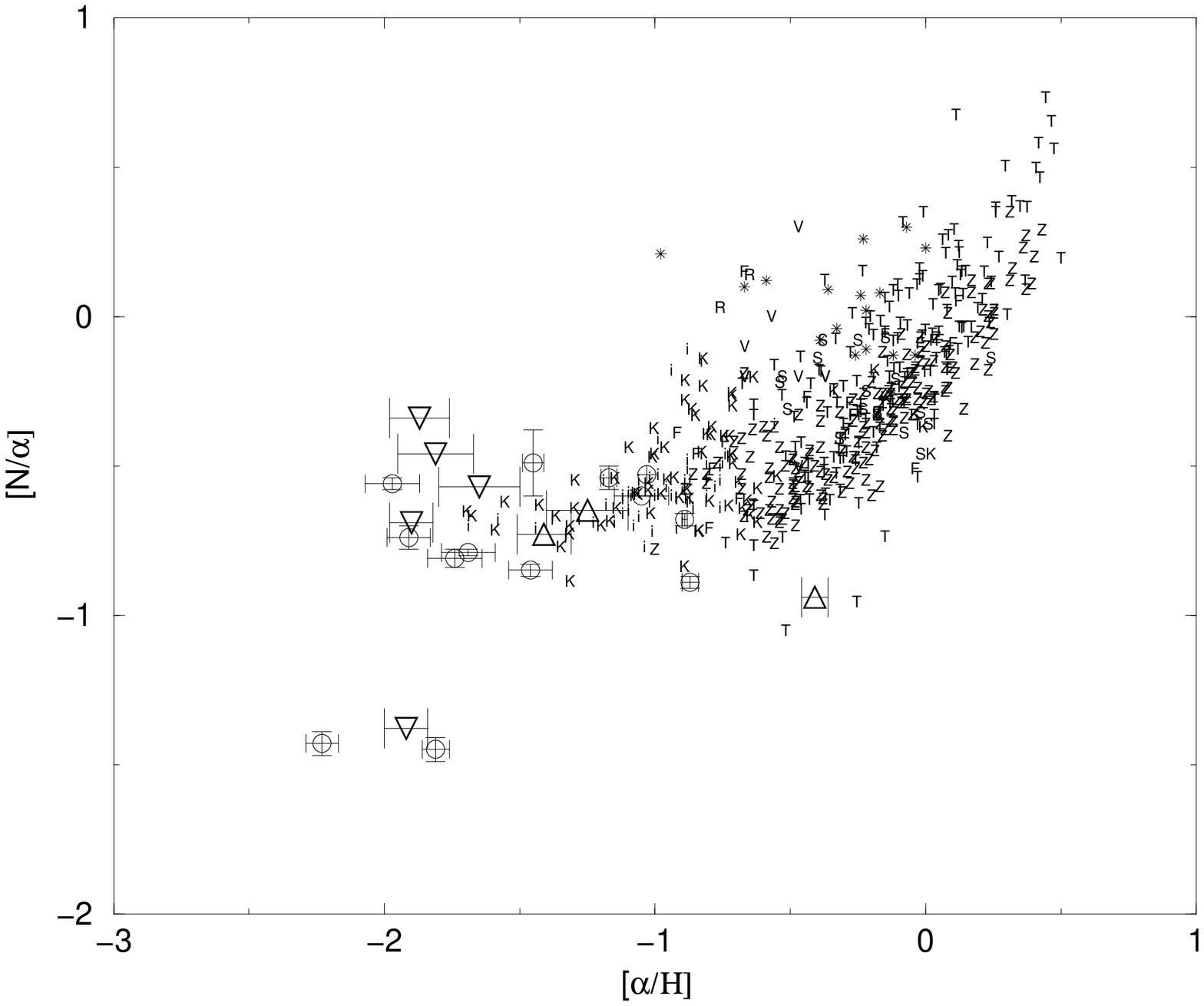}
\caption{[N/$\alpha$] versus [$\alpha$/H] for a large sample of H~II regions and stars compiled by Henry, Edmunds, \& K{\"o}ppen (2000) and Henry \& Worthey (1999) and the damped Lyman alpha systems studied in this paper. Symbols for all but the DLAs are defined in Henry et al. (2000). The DLAs are shown with open circles and error bars. Up triangles represent lower limits, down triangles upper limits. 
}
\label{fig:H1}
\end{center}
\end{figure*}

\section{ANALYSIS AND DISCUSSION}
\label{sec:analysis}

\subsection{What The Data Reveal}

Abundance results from Table~\ref{tab:sum} are presented graphically 
in Figure~\ref{fig:H1}. 
This figure shows the abundance ratios [N/$\alpha$] versus [$\alpha$/H] 
for a large sample of H~II regions and blue compact galaxies compiled by 
Henry \& Worthey (see their Figure~6 along with the caption, which identifies 
the numerous data sources).  To their sample we 
have now added our DLA results, 
which are shown in the figure with open circles, triangles (limits), 
and error bars.

The quantity [$\alpha$/H] in the figure is a gauge of metallicity and 
refers to those heavy elements which are produced 
through subsequent additions 
of alpha nuclei.  In the case of the Henry \& Worthey sample, $\alpha$ 
corresponds to oxygen, while for the objects in our sample, it refers to 
silicon in most cases, and sulfur in a few cases (Table~\ref{tab:sum}).
Metallicity is in turn a measure of the extent to which the observed gas, 
which presumably began as pristine and metal-poor in these objects, has been 
processed through stars, since metal abundances rise as nuclear products from 
evolved stars are added to the gas. Metallicity is directly related to the 
total amount of star formation occurring since the system formed, and 
therefore values increase with time with a slope 
determined by the star formation 
rate and are tempered by the accretion of primordial 
gas from the surrounding IGM.
On the other hand, [N/$\alpha$] measures the differential change in nitrogen 
relative to $\alpha$-elements as processing goes forward. It is sensitive to 
the stellar 
yields for each element as well as the form of the initial mass function, but 
relatively insensitive to the detailed history of the star formation rate.

As we study the global pattern created by the emission line objects in 
Figure~\ref{fig:H1}, we identify a lower-bound envelope 
behind which points are 
concentrated in number densities that fall off with 
increasing distance from the 
envelope.  Referring to the discussion in $\S$~\ref{sec:intro}, 
the plateau in the 
envelope at low metallicity that is displayed by the metal-poor galaxies 
\citep{izotov99,kobulnicky96} is explained by the dominance of primary 
(metal-insensitive) nitrogen production at low metallicity. 
Likewise, the upturn 
at higher metallicities can be attributed to increasing 
contributions from secondary 
(metal-sensitive) production. Currently, the causes of the scatter behind the 
envelope are not well-understood, although several 
theories have been proposed 
\citep{garnett90,pilyugin93}.

Most of the data points in Figure~\ref{fig:H1} 
collectively represent emission 
line regions in many different host galaxies of several different galaxy 
morphologies. Therefore, the fact that we see a pattern 
formed by this diverse 
group of objects suggests that, to first order, 
universal synthesis patterns of 
stars, as communicated to the environment through their IMF-weighted
yields, are more 
important than local factors such as the star 
formation rate, infall, outflow, 
etc. for determining how nitrogen and $\alpha$-element abundances evolve with 
respect to each other. 

Now consider the damped \lya systems plotted in Figure~\ref{fig:H1}.
The majority follow a lower metallicity extension of the 
\nalph\ plateau expressed by the emission line regions, and unlike previous 
studies (e.g.\ L98), we observe minimal scatter in this
group.  The DLA 'plateau' values exhibit an {\it rms} scatter $< 0.2$~dex
which is remarkably small given the systematic 
uncertainties described in $\S$~3.
Furthermore, all of the limits in this sub-sample are consistent with an
[\nalph] value of $-0.75$~dex.
In contrast with the emission line regions, we identify three low
nitrogen damped systems (LN-DLA) which represent a region 
of the plot that is isolated from and discontinuous with the main pattern.
Similar to the major DLA population,
these LN-DLA also exhibit minimal dispersion in their \nalph\ values. 
Therefore, the DLA systems form a bimodal distribution where the majority
lie on the metal-poor plateau and a sub-sample exhibit significantly lower
\nalph\ values.

This bimodality of the DLA \nalph\ measurements was not apparent in previous 
studies.  In part, the difference is our reliance on 
higher quality observations,
but of similar importance is our greater appreciation of the effects of 
photoionization.  Our current sample is small enough, however, that we
do not have complete confidence in a bimodal distribution.
It is worth noting, however, that the new measurements presented in 
\cite{ptt02} lend further support to this description.
In the following, we will investigate the implications 
of the LN-DLA sub-sample
and the impact of a bimodal distribution on various models of N production
and star formation.

\subsection{Model Interpretations} 

Several studies cited in {\S}1 indicate that IMS 
produce the bulk of the nitrogen in the universe. 
Therefore, we expect that the LN-DLA 
arise as the result of a reduced IMS contribution to 
nitrogen buildup. Here, we list four ways that such a reduction could occur.
\begin{enumerate}

\item The LN-DLAs are observed during a transient period occurring within 
250~Myr of a star burst and before the slower evolving IMS have released 
their nitrogen into the interstellar medium of these objects.

\item Nitrogen yields of IMS significantly decline 
with a decrease in metallicity.

\item The observed [N/$\alpha$] reflects the nucleosynthesis characteristics 
of Population~III stars.

\item The IMF in these objects is either truncated at the low mass end or
the slope is significantly flatter than the commonly observed form, 
resulting in a much larger ratio of massive stars to IMS and reducing 
the contribution of IMS to nitrogen levels.

\end{enumerate}
Because the first three of these have been put forth by other authors, 
we only describe them briefly and then turn our attention to our fourth point.
For each option, we examine the scenario under the constraint that it 
naturally reproduces a bimodal distribution of \nalph\ values.  We will find
that this constraint severely challenges the first 
three models and, therefore,
our results favor a top heavy IMF mode of N production in the LN-DLA.

\begin{figure}[ht]
\begin{center}
\includegraphics[height=3.6in, width=2.8in,angle=-90]{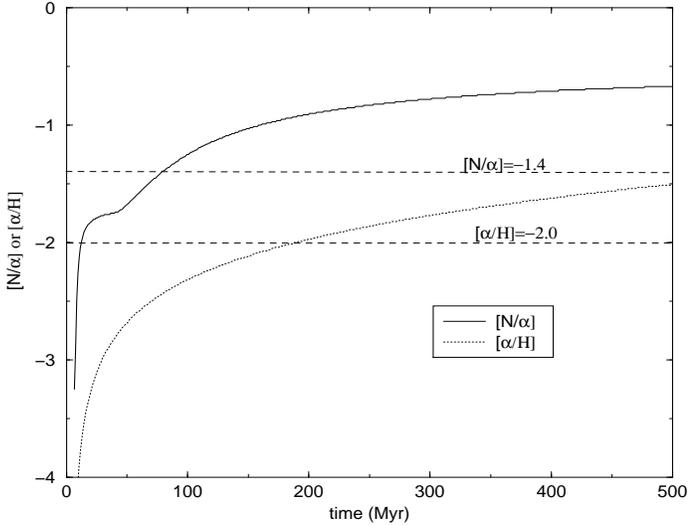}
\caption{
[N/$\alpha$] (solid curve) and [$\alpha$/H] (dotted curve) versus time in Myr 
for model~B of HEK00. The horizontal dashed lines indicate centroid values for 
observed [N/$\alpha$] and [$\alpha$/H] associated with the LN-DLA objects 
displayed in Figure~12.
}
\label{fig:H2}
\end{center}
\end{figure}

The first hypothesis has been considered previously (L98; HEK00; Pettini et
al.\ 2002) and is a reasonable possibility when one 
considers that at $z \sim 3$,
the age of the intervening galaxies must be less than $\approx$2~Gyr, i.e., 
the age of the universe. 
It is possible that we are observing the damped \lya systems during
very early formation times when their 
IMS would not have fully expelled their N products. 
This was shown to be a viable explanation by HEK00, who estimated the time 
delay in IMS nitrogen release to be about 
250~Myr\footnote{Siess, Livio, \& Lattanzio (2002) have recently calculated 
IMS evolution models in which they find that the stellar mass 
threshold for N production 
drops at lower metallicity, thus increasing the timescale for 
its production by IMS even further.}.
The challenge for this scenario is revealed by Figure~\ref{fig:H2} which 
plots [N/$\alpha$] and [$\alpha$/H] versus time in Myr for model~B of HEK00, 
the model which passes nearest the location of the LN-DLAs. 
Overplotted on the theoretical curves are the \nalph\ and \alphH\ values
(horizontal dashed lines) of the LN-DLA sample.
The plots show that the required levels are not reached at the same time:
[N/$\alpha$] reaches a level of $-1.4$ at the same time that 
[$\alpha$/H] is about 0.5 below $-2.0$~dex.  This inconsistency could be 
overcome, however, by fine-tuning the star formation efficiency.
More important is that model~B and its variants predict the gas will spend 
as much time at [\nalph]~$\approx -1.4$
as it will at $-1.2$, $-1.6$~dex, and other low values.  
Therefore, one would expect the LN-DLA to uniformly fill the region below the
higher \nalph\ plateau expressed by the majority of DLA.
The identification of the
three LN-DLA at nearly the same \nalph\ value must be explained as
a coincidence, i.e., this scenario cannot naturally reproduce a bimodal
distribution.

In considering option 2 above, we employ a simple closed-box
model of chemical evolution in order to estimate the ratio of nitrogen to 
oxygen yields needed to explain the observed 
[N/$\alpha$] of $-$1.4 of the LN-DLA. 
Recall that the metallicity in a simple model 
can be expressed as $Z_x=-y_xln\mu$, 
where $Z_x$ is the mass
fraction of element $x$, $y_x$ is the total yield 
of a population of stars characterized by a Salpeter 
IMF per unit of stellar material forever locked up 
in remnants, and $\mu$ is the gas
fraction (Tinsley 1980). From the above observed value 
of [N/$\alpha$] and the solar value for N/O of --0.9 
(Grevesse \& Noels 1996) we infer that $Z_N/Z_O = y_N/y_O=0.005.$ 
HEK00 integrated several sets of published yields over a 
Salpeter IMF, and we can now compare their results for low 
metallicity directly with this observed value. Such a 
comparison shows that current published yield predictions of N/O are about 
two orders of magnitude above the level implied by the 
LN-DLA observations. Their results 
appear to rule out option~2 above, although published IMS yield calculations 
so far do not extend down to metallicities below 0.001. 
Furthermore, there is no indication from published IMS nitrogen yields that 
they are highly sensitive to metallicity (van~den~Hoek \& Groenewegen 1997; 
Marigo 2001).  Finally, it would be very difficult to explain the bimodality
of [\nalph] values at the same [\alphH] metallicity in this scenario.

For option~3 we compare in a similar way the same 
simple model results to predicted yield ratios. 
Heger \& Woosley (2002) have calculated models of massive Pop~III stars 
ranging in mass from 140 to 260~M$_{\sun}$ and predicted yields for $^{14}$N 
and $^{16}$O. 
Simply by inspecting their results we can see that values of N/O resulting 
from their models range from about 10$^{-6}$ in the least massive stars 
to 10$^{-7}$ in the most massive ones. In addition, Woosley \& Weaver (1995) 
predict N and O production by zero metallicity
stars between 12-40~M$_{\sun}$ and find values 
of N/O of a few times 10$^{-4}$, where again we have used the 
HEK00 results for yields integrated over a 
Salpeter IMF for comparison. These values 
are lower than necessary to explain LN-DLAs. However, in their Population~III 
stellar models Umeda, Nomoto, \& Nakamura (2000) predict $^{14}$N/$^{24}$Mg 
values of roughly 1/4 solar, 
which they say implies a large contribution of Population~III stars to the 
early nitrogen buildup in their yield ratios which are consistent with these. 
In summary, while the relevance of Population~III stars is presently unclear,
these stars could ultimately prove important for 
the issue of early nitrogen production.
Furthermore, if one could introduce a Pop~III model which reproduced the 
observed LN-DLA population, it could reasonably account for the bimodal
distribution where the higher \nalph\ DLA represent a second generation
of star formation.

For the remainder of the discussion, we explore the fourth hypothesis: 
star formation is either characterized by (1)~a top-heavy
initial mass function at early times and/or low metallicities, 
such that there is a relative deficiency of IMS; 
or (2)~a stellar population that is truncated below 
a certain mass threshold. In either case, the observed [N/$\alpha$] 
in the LN-DLAs is dominated by the yield patterns of 
massive stars.

\begin{figure*}[ht]
\begin{center}
\includegraphics[height=3.6in, width=3.3in,angle=-90]{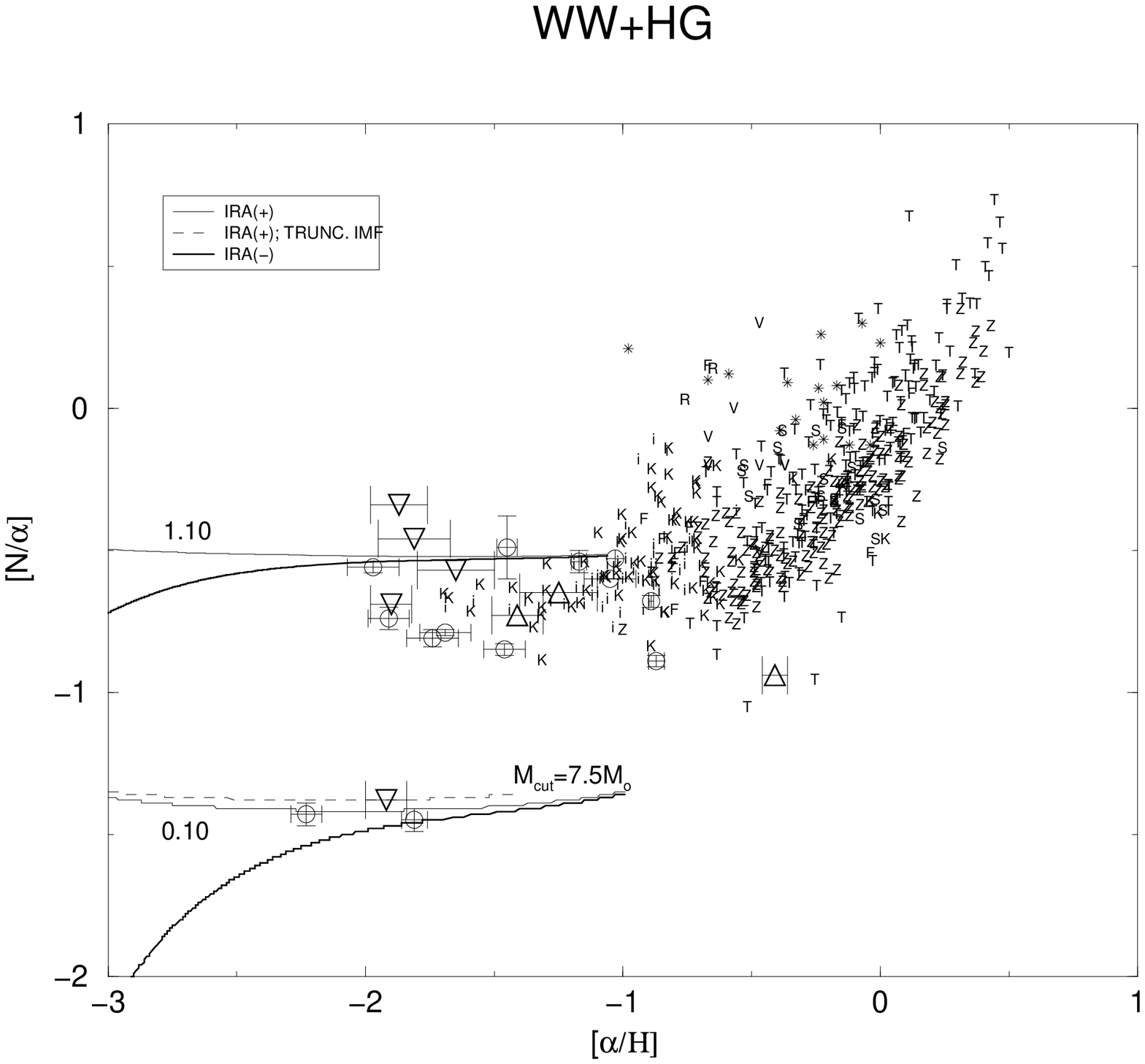}
\includegraphics[height=3.6in, width=3.3in,angle=-90]{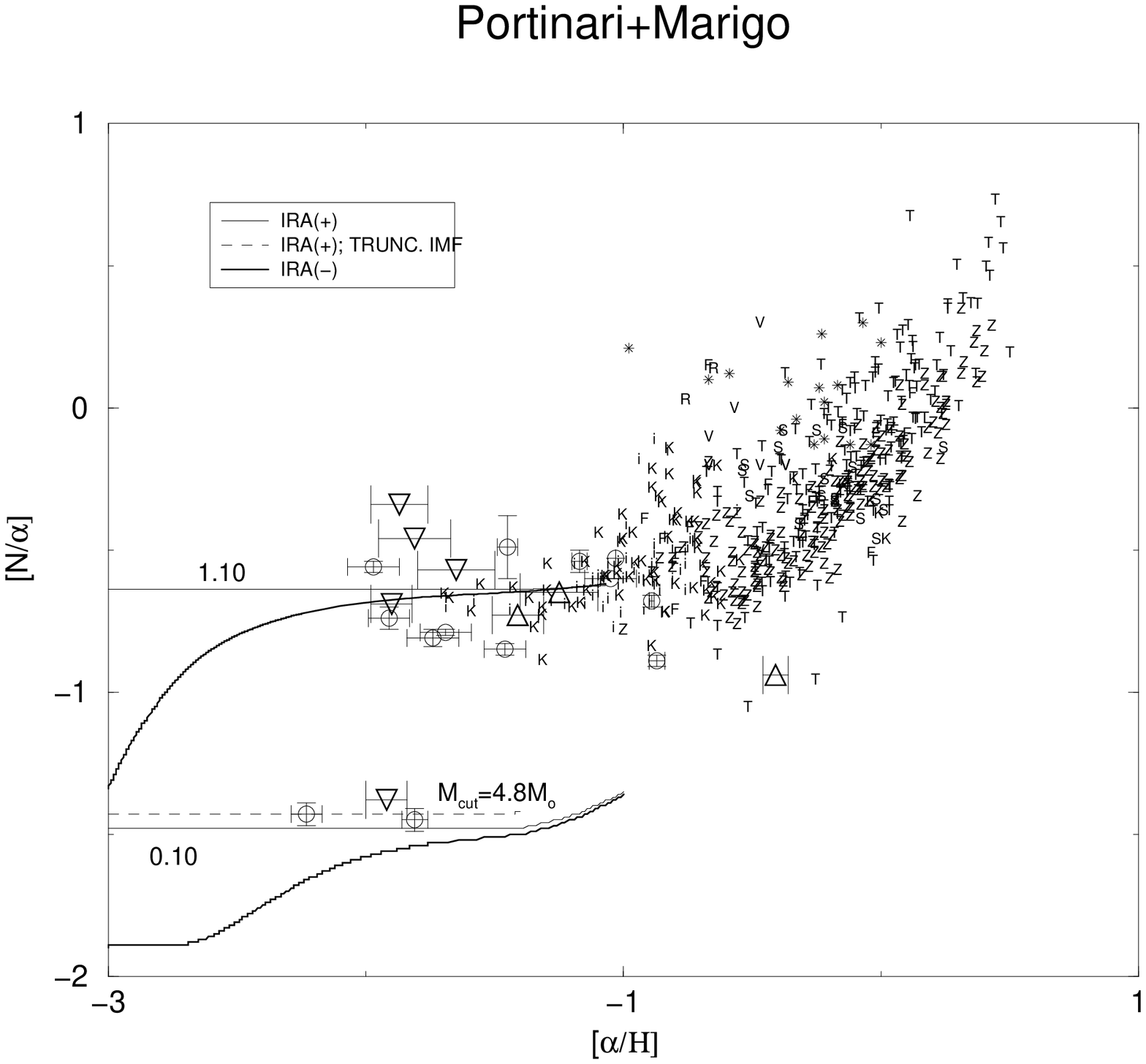}
\caption{
(a) Model results for set 1 stellar yields displayed with data points of 
Figure~12. Light solid curves represent results for the two different IMF slopes 
indicated and when instantaneous recycling is assumed (IRA+). Heavy solid lines 
show tracks for same models except that the instantaneous recycling assumption 
is relaxed (IRA-). The dashed line shows the model result when the IMF is 
truncated below 7.50~M$_{\sun}$ and instantaneous recycling is assumed. (b) Same 
as (a) but for set 2 stellar yields.
}
\label{fig:H3}
\end{center}
\end{figure*}

To test this hypothesis, we employ the numerical code 
of HEK00 and calculate two types of chemical evolution 
models, both of which assume a closed system, i.e. no 
matter exchange with the environment. In the first set 
of models we ignore evolutionary time lags among stars 
of different progenitor masses (instantaneous recycling 
approximation), while in the second set we relax this 
assumption. Otherwise the model types are identical. The 
numerical code we employed is described in detail in HEK00. 
The only alterations we made relate to the star formation rate 
(eq.~18 in HEK00), where we used 1.5 rather than 2.0 for the 
exponent and set the efficiency factor to a constant value 
of 0.003 throughout.

We tested two sets of yields comprising the following combinations: 
(1)~Woosley \& Weaver (1995; massive stars) and van~den~Hoek \& Groenewegen
(1997; IMS); 
and (2)~Portinari et al. (1998; massive stars) and Marigo (2001; IMS). 
Our calculations, therefore, accounted for the effects of progenitor mass 
and metallicity on the yields. The IMF was taken to be a power law form, 
the index of which we varied so as to alter the relative mix of massive 
stars and IMS. For the models which assumed instantaneous 
recycling, we calculated an additional 
model subset for each yield set using a Salpeter IMF (index=1.35) 
and experimented with different lower mass threshold 
values until finding one which satisfactorily reproduced 
the observed [N/$\alpha$] value. Besides the choice of yield set, the 
only parameters 
allowed to change in our models were the slope and 
lower mass threshold of the IMF.

The model results assuming instantaneous recycling 
are shown with solid lines in Figures~\ref{fig:H3}a,b, where each graph 
represents models associated with one yield set 
(see figure legends for line-type definitions). Lines are labeled according 
to IMF slope or cutoff mass\footnote{Note that because we 
are only interested in the 
results at low metallicity, all models were stopped when [$\alpha$/H]$=-1$.}. 
Because of the instantaneous recycling assumption, 
no time lag exists between IMS and massive star 
release of N into the interstellar medium, and thus the tracks 
for these models are horizontal lines. 
In Figure~\ref{fig:H3}a, where we employed the 
set~(1) yields, the positions of the three LN-DLAs are consistent with a 
system whose stellar population possessed a top-heavy IMF with a slope 
of roughly 0.10 or with a Salpeter IMF but truncated below 7.5~M$_{\sun}$. 
In Figure~\ref{fig:H3}b, using set~(2) yields, the LN-DLAs are again 
consistent with an IMF slope of 0.10 or a mass cutoff of 4.8~M$_{\sun}$. 
Note that for each yield set a slope of 1.10 produces reasonable fits 
to the plateau region for both yield 
sets\footnote{The difference here between 
1.10 and 1.35 (Salpeter slope) is not of immediate concern, since 
uncertainties in the yields themselves are roughly a factor of two.}. 
Finally, we point out that [N/$\alpha$] is quite sensitive to the value 
of the mass cutoff in both cases, since it falls within 
the mass range where most IMS N is produced in both cases.

Results for models without the instantaneous recycling approximation
are shown with bold lines. The IMF slope in each case is identical to 
the model track with which the bold line merges close to [$\alpha$/H]=--1. 
The gradual rise in [N/$\alpha$] with 
metallicity for these models is simply 
due to the delayed contribution of IMS 
to the N buildup. Clearly, these models 
are still capable of reproducing DLA observations 
for both the plateau and LN-DLA objects.

Thus, our closed box models, calculated using two different 
sets of theoretical stellar yields, appear to be consistent with the idea 
that LN-DLAs 
are systems which have evolved with either top-heavy or truncated IMFs. 
We find a representative IMF slope of 0.10 or a mass cutoff of 4.8
M$_{\sun}$ with yield set~(1) and 7.5~M$_{\sun}$ with yield set~(2). 
This conclusion holds regardless of whether instantaneous 
recycling is assumed. 
The reason that the altered IMF slope or the truncation can 
explain the LN-DLAs is simply that 
the relative contribution of IMS to nitrogen evolution is 
significantly reduced, leaving the massive stars to 
dominate the production of both nitrogen and oxygen.

The top-heavy IMF hypothesis, we believe, offers a viable alternative 
to ones already suggested by others and briefly discussed above. 
Furthermore, this scenario may also explain the bimodal distribution observed
for the DLA measurements (see below).
{\it Indeed, the LN-DLAs may offer the first evidence that the mass 
spectra of some early stellar populations were much different 
than we observe today.} 

There is currently no convincing evidence for the existence, 
past or present, of an IMF which differs markedly from those 
derived from observations in local systems such as the form 
presented by Scalo (1986). Instead, most evidence points 
to the IMF being invariant (e.g.\ Kroupa 2001) in space and time.
However, simple arguments based upon a Jeans mass analysis imply that at low 
metallicities, star formation should favor higher mass objects. This factor alone alerts us to a 
potential top-heavy IMF (THIMF) at early times.

Silk (1998) discusses several observations 
which may suggest a THIMF at very early epochs,
although he is very careful to note that they are all
circumstantial and in no way clinches the case for a THIMF.
Theoretical support for a THIMF comes from the models of 
Nakamura \& Umemura (2001; 2002), who study star formation in extremely 
metal-deficient protogalactic environments and predict a resulting bimodal 
IMF with peaks around 1 and 100~M$_{\sun}$.

At present, we favor the THIMF interpretation because it accurately reproduces
the \nalph\ values of the LN-DLA and because it may be consistent with a 
bimodal distribution.
The latter point would hold if a primordial THIMF abruptly changed 
over to the form we observe today after a certain threshold was passed. 
\cite{bromm01} have experimented with 
numerical simulations of the collapse of pre-enriched 
primordial objects and set a value of $Z_{crit}=5 \times 10^{-4} Z_{\odot}$ 
as the metallicity threshold when this changeover occurs 
(see also Abel, Bryan, \& Norman 2000). 
In the end it seems likely that measurements of integrated light 
from early systems will prove useful in deciding between the THIMF option 
and the other possibilities which are all consistent with a normal stellar 
mass spectrum.

\section{SUMMARY AND CONCLUDING REMARKS}

In this paper we presented \nalph\ vs.\ \alphH\ measurements
of a moderate sample of damped \lya systems and identified a possible
bimodality in the \nalph\ values
at the lowest metallicity.  While the majority of low metallicity DLA
have \nalph\ values consistent with low metallicity emission-line 
regions in the local universe, a sub-sample exhibits significantly
lower \nalph\ values (the LN-DLA).
Unlike previous works, we found minimal scatter among the members of
each of the sub-samples and therefore argue that the DLA measurements form a
bimodal distribution.
We considered several scenarios which could account for these observations:
(1) these systems are too young ($<250$~Myr) for their intermediate-mass 
stars to have produced N;
(2) a reduced production of N in intermediate-mass stars at low metallicity;
(3) Population III nucleosynthesis;  and
(4) chemical evolution with a top-heavy IMF.
Currently, we do not favor the first three scenarios.  If the LN-DLA are
to be explained through the young age of the absorbing galaxy, then 
one would expect the LN-DLA to fill the parameter space beneath the plateau
instead of comprising a bimodal distribution.
Regarding the second option, it would be difficult to explain why DLA with
the same metallicity exhibit such different \nalph\ values.
Finally, the third avenue is promising but current models 
\citep[e.g.][]{heger02} of Pop~III nucleosynthesis
appear to underpredict the \nalph\ levels observed in the damped systems.

Therefore, we favor a scenario where
the LN-DLA were enriched with an initial starburst with suppressed 
formation of intermediate-mass stars.
We found that a starburst characterized by a power-law IMF with slope 0.10 
or a mass cut of $\approx 5-8 \msol$ would successfully reproduce 
the observed LN-DLA values.
This scenario allows for the bimodal DLA distribution of \nalph\
without requiring that we have observed the damped systems at a special moment
in time.  
Furthermore, 
an initial burst of star
formation in a low metallicity environment may suppress the formation of
lower mass stars \citep[e.g.][]{abel00,bromm01} and naturally lend to a
bimodal distribution.

Before concluding, we wish to briefly
comment on correlations between \nalph\ and
other physical properties of the damped \lya systems.
In terms of kinematic characteristics, 
two of the LN-DLA systems (Q1946+75, z=2.8; Q2348--14, z=2.3) have
among the simplest kinematics of any damped \lya system.  
This may suggest these systems have particularly small mass and/or
relatively quiescent star formation.  
We note, however, that several other systems with higher \nalph\ ratios
also exhibit very simple kinematics.
The only other trend we can identify is an increase in \nalph\ with increasing
$z_{abs}$ which is contrary to expectation.
Perhaps this absence of correlation is further evidence that the production
of N is largely independent of environment and star formation history as
suggested by the uniformity of the emission line sample.

\acknowledgements

The authors wish to recognize and acknowledge the very significant cultural
role and reverence that the summit of Mauna Kea has always had within the
indigenous Hawaiian community.  We are most fortunate to have the
opportunity to conduct observations from this mountain.
We thank the Keck staff who were instrumental in acquiring this data.  
We also thank Max Pettini and Sara Ellison for their helpful comments.
We also acknowledge the construction of the MAKEE software package
by Tom Barlow.
Finally, we thank Gary Ferland and CLOUDY associates for the CLOUDY
software package.
This work was partially supported by NASA through a Hubble Fellowship
grant HF-01142.01-A awarded by STScI to JXP.
RBCH was partially supported by NSF grant AST-9819123.
The work of DT, NS, JMO, and DK was funded in part by 
grant NASA funds G-NASA/NAG5-3237 and NAG5-9224, and by
NSF grant AST-9900842.
AMW was partially supported by NSF grant AST 0071257.



\end{document}